\documentclass[10pt, conference, compsocconf]{IEEEtran}

\ifCLASSINFOpdf
\else
\fi
\usepackage{multirow}
\usepackage{graphicx}
\usepackage{rotating}

\renewcommand\baselinestretch{0.95}

\begin{document}
%
\title{Adaptive Performance Optimization under Power Constraint in\\
 Multi-thread Applications with Diverse Scalability}

\author{\IEEEauthorblockN{Stefano Conoci, Pierangelo Di Sanzo, Bruno Ciciani}
\IEEEauthorblockA{DIAG - Sapienza University of Rome\\
Email: \{conoci.1483662@studenti.uniroma1.it, \\disanzo@dis.uniroma1.it, ciciani@dis.uniroma1.it\}}
\and
\IEEEauthorblockN{Francesco Quaglia}
\IEEEauthorblockA{DICII - University of Rome Tor Vergata\\
Email: francesco.quaglia@uniroma2.it}
}


%


\maketitle

\begin{abstract}

In modern data centers, energy usage represents one of the major factors affecting operational costs. Power capping is a technique  that limits the power consumption of individual systems, which allows reducing the overall power demand at both cluster and data center levels.
However, literature power capping approaches do not fit well the nature of important applications based on first-class multi-thread technology. For these applications
 performance may not grow linearly as a function of the thread-level parallelism because of the need for thread synchronization while accessing shared resources---such as shared data. In this paper we consider the problem of maximizing the application performance under a power cap by dynamically tuning the thread-level parallelism and the power state of the CPU-cores. Based on experimental observations, we design an adaptive technique that selects in linear time the optimal combination of thread-level parallelism and CPU-core power state for the specific workload profile of the multi-threaded application. We evaluate our proposal by relying on different benchmarks, configured to use different thread synchronization methods, and compare its effectiveness to different state-of-the-art techniques.

\end{abstract}


%
\IEEEpeerreviewmaketitle

\section{Introduction}
\label{introduction}

Multi-core architectures are nowadays dominating the market. Also, thanks to the support they offer for sharing memory among CPU-cores, they have become the mainstream reference hardware for applications based on the first-class multi-thread technology.
%
%
%
On the downside, powering many-core machines implies high energy delivery to each single multi-core server. Therefore,
over the last years, energy and power consumption raised up as a core concern to cope with, especially in (large) data centers.

Such concern led manufacturers to introduce hardware mechanisms oriented to improve energy efficiency in operational contexts. These include Dynamic Voltage and Frequency Scaling (DVFS), which allows lowering the voltage and the frequency (hence the power consumption) of a processor/core in a controlled manner, and Clock Gating, which disables some processor/core circuitry during idle periods. Contextually, today's Operating Systems offer power management tools---like Linux CPUFreq Governor \cite{Palli_06_ondemand}---which expose to the user code interfaces to dynamically change the power state of cores via DVFS, thus allowing to tune the performance of cores and their power demand according to the need of specific applications/workload.


In this context, one interesting challenge is the one of controlling the power demand of an application in order to keep it below a given threshold, also known as the {\em power cap}.
However, an even more interesting challenge is the one of ensuring that an application runs at maximum performance under a given power cap. Such an achievement, in addition to performance benefits, would also improve the application efficiency in terms of energy per task.

Various power capping techniques for multi-core servers have been proposed.
As for the specific case of multi-thread applications, the problem of regulating the number of threads and the core frequency to control the balance between performance and power consumption has been originally considered in \cite{Reda:2012}, and subsequently in \cite{Zhang_2016}. The main drawback of these approaches is that the tuning strategies they rely on do not account for complex and dynamic effects on performance that may be caused by thread contention on hardware resources and/or shared data.
In more details, when multiple threads concurrently run on different CPU-cores, the presence of shared hardware resources---such as memory interconnections and cache levels---leads them to contend for their utilization. This impacts both performance and the power consumption profile of the application. Also, in common multi-thread applications that are not disjoint-access parallel, threads share data whose accesses may require synchronization.
This still affects performance and the power consumption profile, also depending on the specific synchronization mechanisms (either speculative or not) that are employed by the application code. An additional factor of complexity in the presence of synchronization is that the speed-up achieved by running the application with different number of threads may change depending on the workload profile, which in turn can be dynamic by its own.  Also, the speed-up can be non-linear as a function of the number of threads, depending on the workload profile, as well as the underlying hardware settings. Specifically, performance can even decrease when increasing the level of parallelism. This indicates that synchronization costs, including the energy spent while performing synchronization operations, can show complex profiles to deal with.


Overall, to select the right combination of thread parallelism and core power state, which ensures the best performance under a power cap, it looks mandatory to take into account the (possible) limited scalability of a multi-thread application, just like it manifests at run-time due to actual synchronization dynamics. Further, it is mandatory to react to variations of the workload profile.

To cope with this problem, we present an adaptive technique that uses a novel on-line exploration-based tuning strategy. We devised our technique exploiting empirical observations of the effects on both performance and power consumption associated with the combined variation of thread-level parallelism and core power state. Specifically, by the results of experiments we conducted with different multi-thread benchmarks characterized by non-negligible incidence of synchronization ---e.g. because of thread contention while accessing share data---we highlight that their scalability
%
%
is not affected by the variation of the power state of the CPU-cores. Based on this, we defined an optimized tuning strategy where the exploration moves along specific directions that depend on the power cap value and on the intrinsic scalability of the application. Remarkably, we prove that the proposed technique finds in linear the optimal configuration of concurrent threads and CPU frequency/voltage, i.e. the configuration that provides the highest performance among the configurations with power consumption lower than the power cap. Also, we present a refinement of our technique that exploits continues fluctuations between configurations---in terms of thread-level parallelism and core power state---to further improve the application performance and reduce the possibility/incidence of power cap violations.

We demonstrate the advantages of our proposal via an experimental study based on various application contexts, including various benchmarks that use different thread synchronization methods. This allows us to robustly assess our technique via disparate test cases where contention among threads affects the application scalability in significantly different ways.

The remainder of this article is structured as follows. In Section \ref{related} we discuss related works. Section \ref{preliminary_study} defines our target problem and presents the results of the preliminary analysis. Section \ref{technique} illustrates the proposed optimization technique, proves that the selected configuration is optimal and analyzes the time complexity of the exploration procedure.
Section \ref{experimental} describes the most relevant implementation details and presents the experimental results.

\section{Related Work}
\label{related}

A work specifically focused on optimizing the energy demand at application level is presented in \cite{Reda:2012}. The proposed technique, called Pack and Cap, aims at selecting the best configuration, in terms of number of cores to be assigned to an application and the related core frequency, which ensures a given power cap for multi-thread workloads. Based on experimental measurements of the performance and power consumption obtained running benchmarks from the Parsec suite, the authors conclude that the configuration that provides the highest performance at a given level of power consumption always assigns to the application the highest possible number of cores. However, as extensively shown in the following of this article, this selection strategy is not optimal for general multi-threaded applications with less than linear scalability.
The work in \cite{Zhang_2016} considers the problem of maximizing performance under a power cap while also taking into account the effects of contention. The solution defines an ordered set of power knobs that are progressively tuned by performing a binary search on the respective domain, selecting the setting that provides the highest performance for the considered power knob while operating within the power cap. In particular, the solution first selects the optimal number of cores that should be assigned to an application while running at the slowest available frequency/voltage, then selects the optimal CPU \textit{P-State} setting for the previously selected number of assigned cores. Therefore, by tuning the power knobs independently, it does not consider the changing energy/performance trade-offs at different levels of parallelism for the specific workload. As an example, if an application shows a limited speed-up when increasing the number of cores, the solution would still pick the highest value that provides a power consumption within the cap, even if the same power budget could provide higher performance if spent to further increase the frequency of a lower number of cores.

Other works in literature investigate the problem of improving application performance under power constraint considering different power management variables. FastCap \cite{Liu2016} defines an approach for optimizing performance under a system-wide power cap considering both CPU and memory DVFS. It defines a non-linear optimization problem solved through a queuing model that considers the interaction between CPU-cores and memory banks communicating over a shared bus. Unfortunately, memory DVFS has only been proposed recently \cite{Deng2012} \cite{David2011} and is not yet available in commercial systems.
Kanduri et al. propose approximation as another knob that can be used in power capping, combined with DVFS and Clock Gating, to define a trade-off between performance and accuracy of the results \cite{Kanduri2016}. However, in order to dynamically switch between different levels of accuracy, it requires multiple implementations of the
same application. PPEP \cite{Su2014} is an online prediction framework that, based on hardware performance events and on-chip temperature measurements, estimates the performance, power consumption and energy efficiency for each different CPU \textit{P-state}. Therefore, it allows the definition of a power capping technique that can meet power targets in a single step without requiring any exploration. However, it does not consider the possibility of altering the number of cores assigned to an application, thus it would provide sub-optimal performance for multi-thread applications showing less than linear scalability.

\section{Problem Statement and Preliminary Analysis}
\label{preliminary_study}

As discussed, in our study we consider the problem of adaptively tuning the system configuration to ensure the highest application performance under a power cap. We consider two tuning parameters, the number of concurrent threads and the cores power state. We focus on the general scenario of multi-thread applications executed on a working-thread pool (e.g. multithreaded web/application servers) whose size can be tuned at run-time. However, we should note that the proposed technique is orthogonal with respect to the chosen thread regulation mechanism.
Also, we assume that the power state of cores can be changed, affecting both power consumption and performance. In practice, this is what happens when changing the so-called \textit{P-state} in modern multi-core processors, which determines a variation of the core voltage and frequency, thus modifying both the power consumption and the instruction processing speed. We adhere to the notation of ACPI standard, which establishes that \textit{P0} denotes the core state with maximum power and performance, and \textit{P1}, \textit{P2}, ... progressively identify states with less power and performance. Also, we consider the core idle state (\textit{C-state}), where \textit{C0} denotes the full operating core state, and \textit{C1}, \textit{C2}, ..., progressively identify lower power states where the core is idle, i.e. it does not execute instructions. A core can transit from \textit{C0} to a deeper \textit{C-state} when it has no instruction to execute. Hence, when the number of running threads goes below the number of available cores, unused cores can transit to low power states, thus reducing the total power consumption.

\begin{figure}
\centering
\includegraphics[width=.4\textwidth]{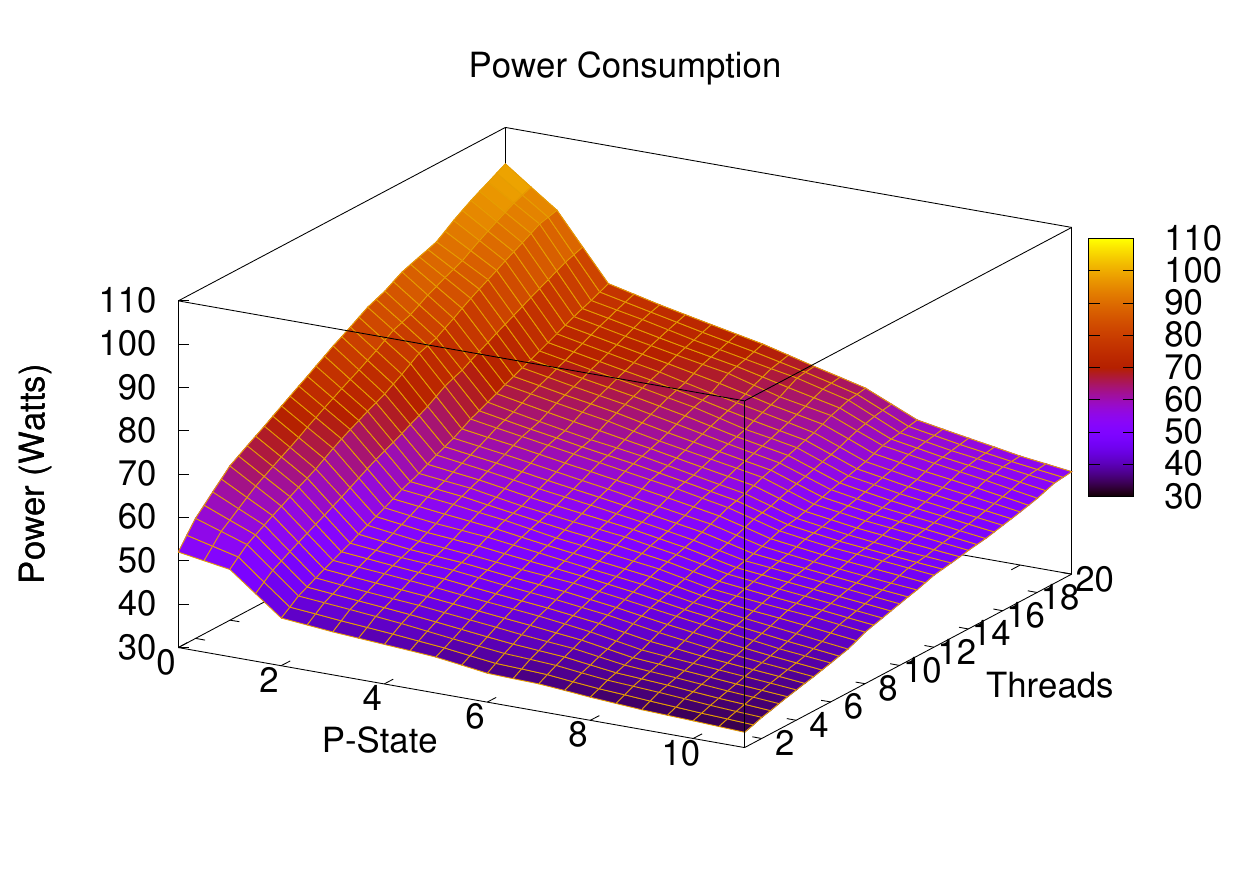}
\vspace*{-0.8cm}
\caption{Throughput vs. Number of Concurrent Threads and \textit{P-state}}
\label{intruder_power_analysis}
\end{figure}

To provide the reader with real data demonstrating the effects on power consumption associated with the variation of \textit{P-state} and the number of concurrent threads, we show in Figure \ref{intruder_power_analysis} the results of an experiment where we run the multi-thread Intruder benchmark from the STAMP suite~\cite{stamp} for Transactional Memory systems \cite{ShavitSTM}.
Intruder emulates a signature-based network intrusion detection system where network packets are processed in parallel by concurrent threads. We executed different runs while changing \textit{P-state} and  the number of concurrent threads on top of a machine with two Intel Xeon E5, 20 physical cores total, 256 ECC DDR4 memory, with core clock frequency ranging from 1.2 GHz (whose \textit{P-state} is denoted as P-11) to 2.2 GHz (denoted as P-1), and TurboBoost from 2.2 GHz to 3.1 GHz (denoted as P-0). Since we focus on the effects of the joint  variation of core power state and thread parallelism, we consider power consumption data related to the CPU and memory subsystems, which we collected via Intel RAPL interface \cite{Intel6437}. The plot shows the power consumption as a function of the couple $(p, t)$, where $p$ is \textit{P-state} and $t$ is the number of concurrent threads. The results clearly outline that the power consumption grows while incrementing either the first or the second variable. Given a power cap value, if $\{(p, t)\}$ is the set of all possible configurations, we denote as ${\{(p, t)\}}_{ac}\subseteq\{(p, t)\}$ the subset of all acceptable configurations, that is the configurations for which the power cap is not violated. Formally, it is the subset such that $pwr(p, t)\leq C$, where $pwr(p, t)$ is the power consumption with configuration $(p, t)$ and $C$ is the power cap value. Since the function $pwr(p, t)$ monotonically increases with respect to both $p$ and $t$, the subsets of acceptable and unacceptable configurations are separated by a frontier, as shown in figure \ref{intruder_exploration}. 

Our goal is to find the configuration ${(p, t)^* \in {\{(p, t)\}}_{ac}}$ for which the performance of the application is maximized. Without loss of generality, we consider the application throughput as a performance metric. In any case, with our approach also other metrics, depending on the specific application, could be used, such as the application runtime or the operation response time. We denote as $thr(p, t)$ the application throughput for configuration $(p, t)$.

In multi-thread applications, the variation of the application throughput as a function on $t$ plays a key role when finding the best configuration. Due to hardware and data contention phenomena, the profile of the application throughput curve is generally characterized by two parts, i.e.  an initial ascending part, where the throughput increases while increasing $t$, followed by a descending part, where the throughput decreases while increasing $t$. However, we note that in the case of high contention the initial ascending part may not exist (i.e. the throughput always decreases when  increasing $t$). Conversely, in the case of low contention the throughput may never decrease. 

\begin{figure*}
\centering
\includegraphics[width=.32\textwidth]{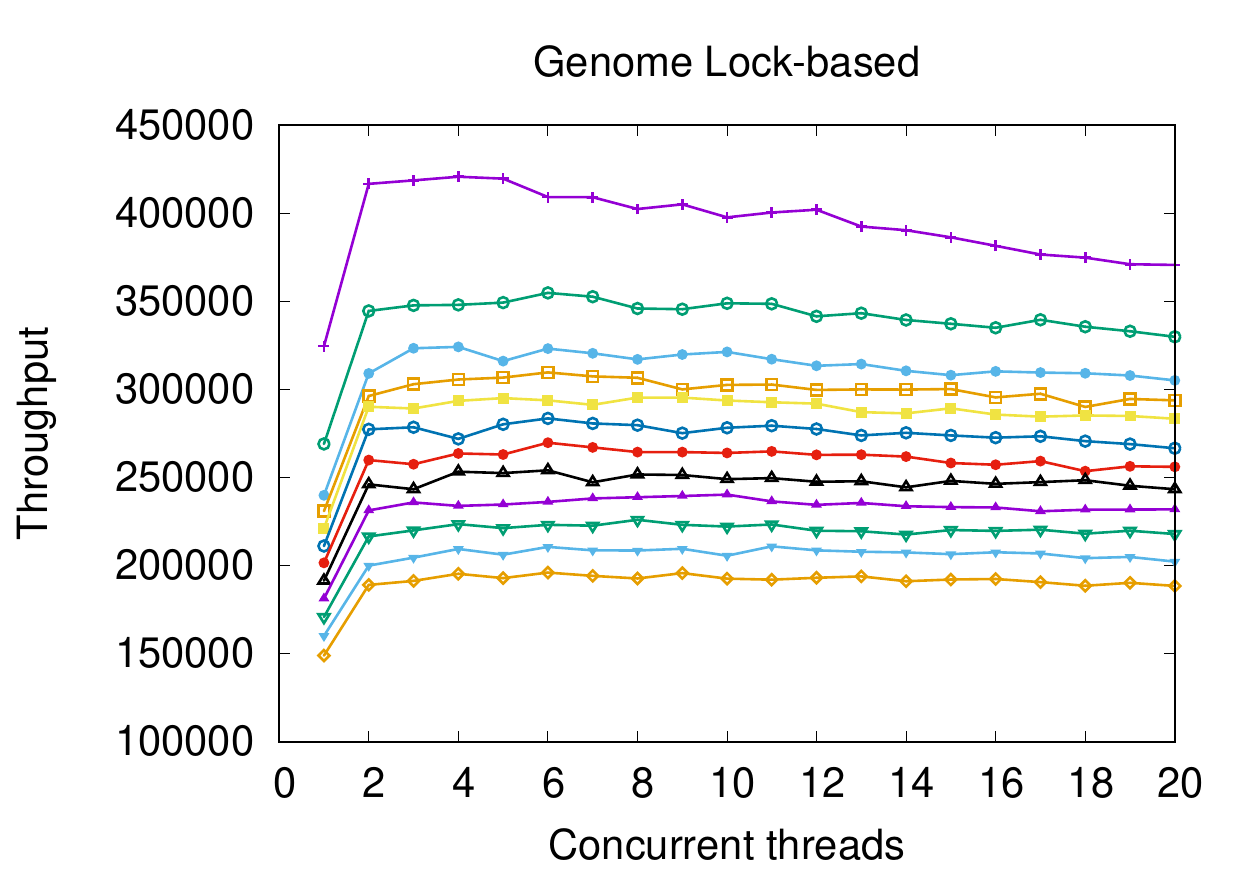}
\includegraphics[width=.32\textwidth]{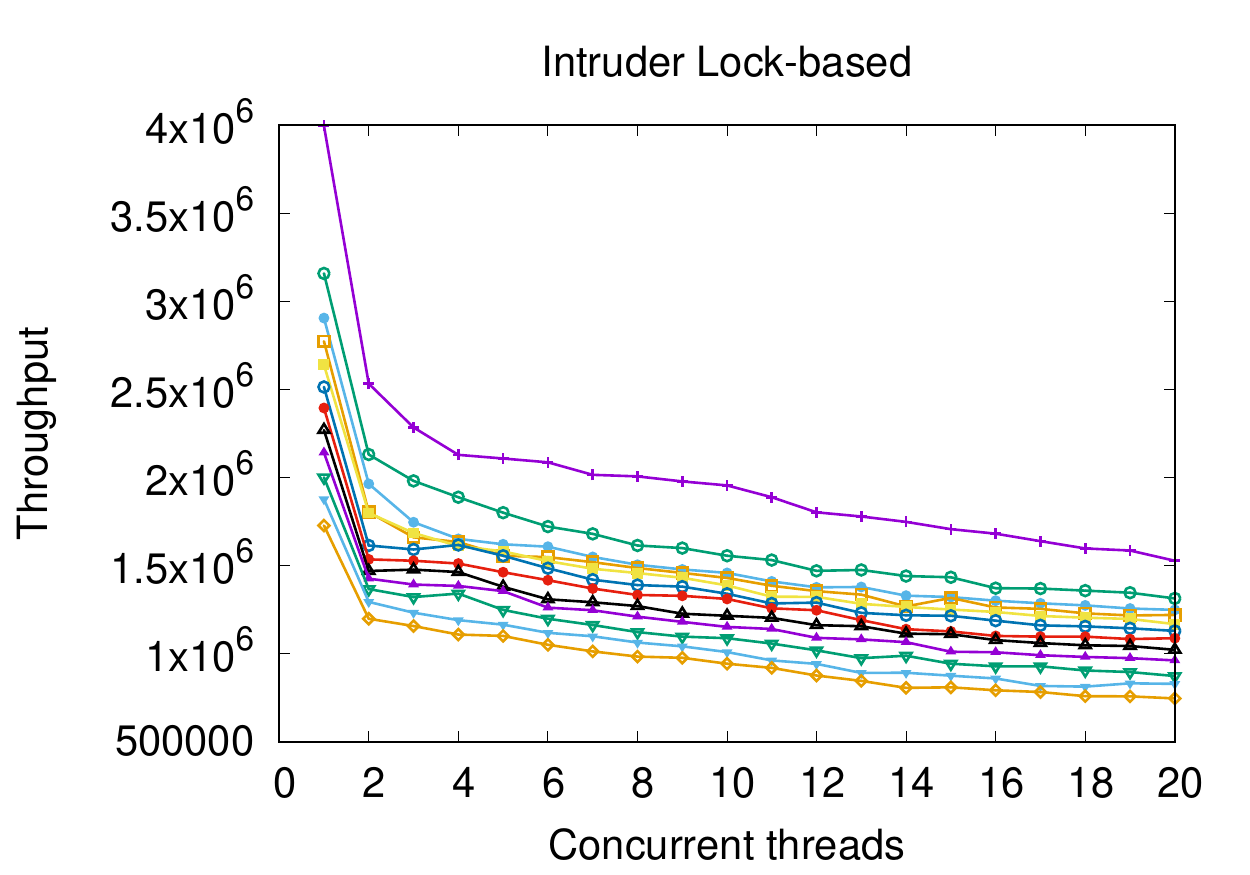}
\includegraphics[width=.32\textwidth]{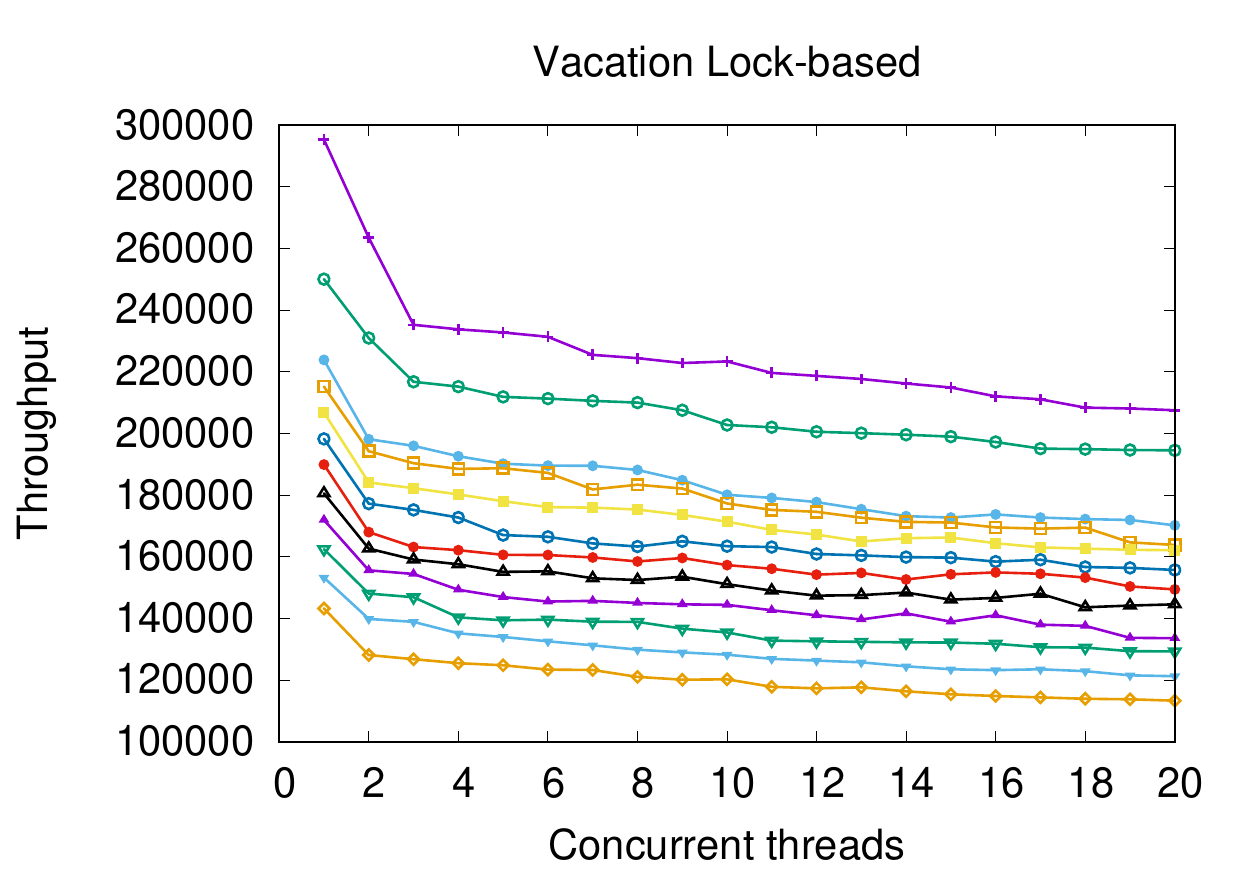}
\includegraphics[width=.32\textwidth]{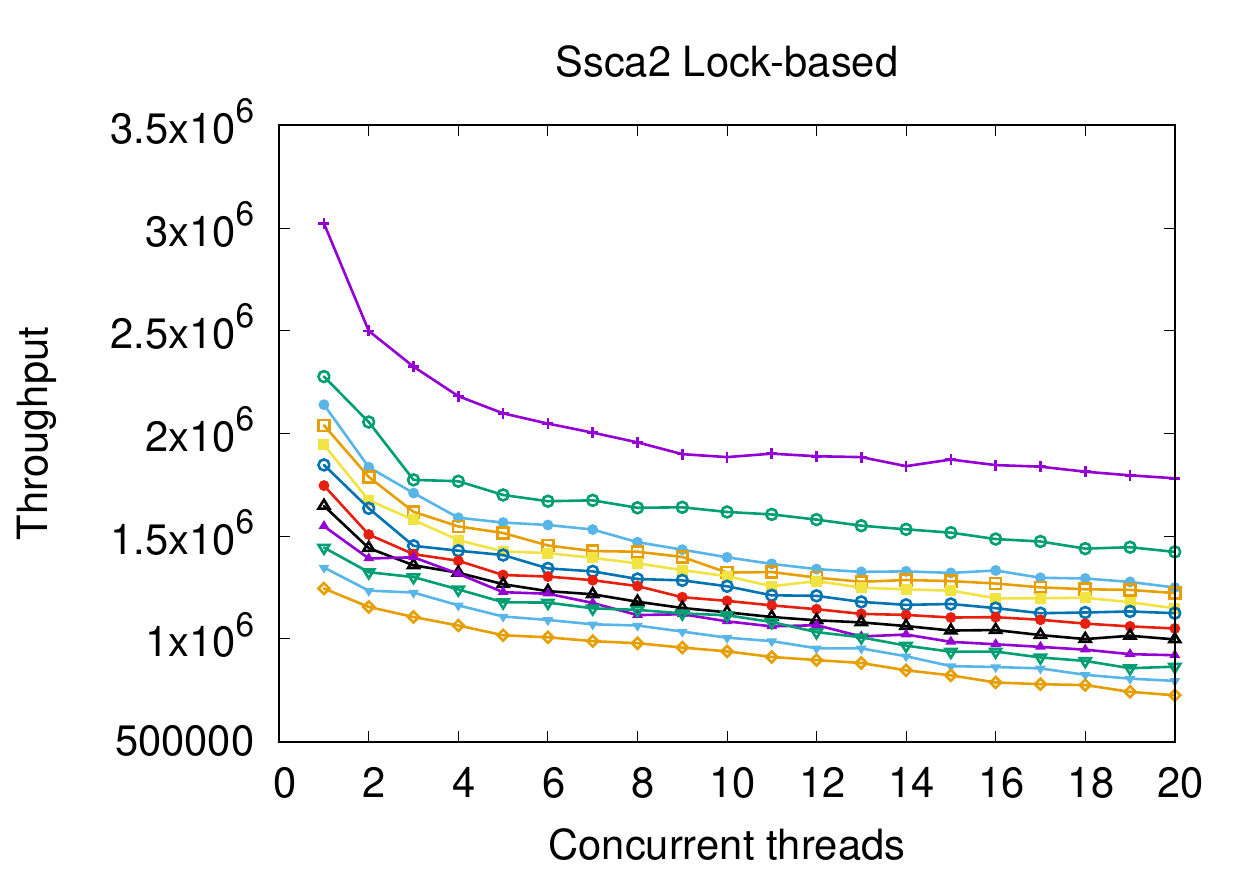}
\includegraphics[width=.32\textwidth]{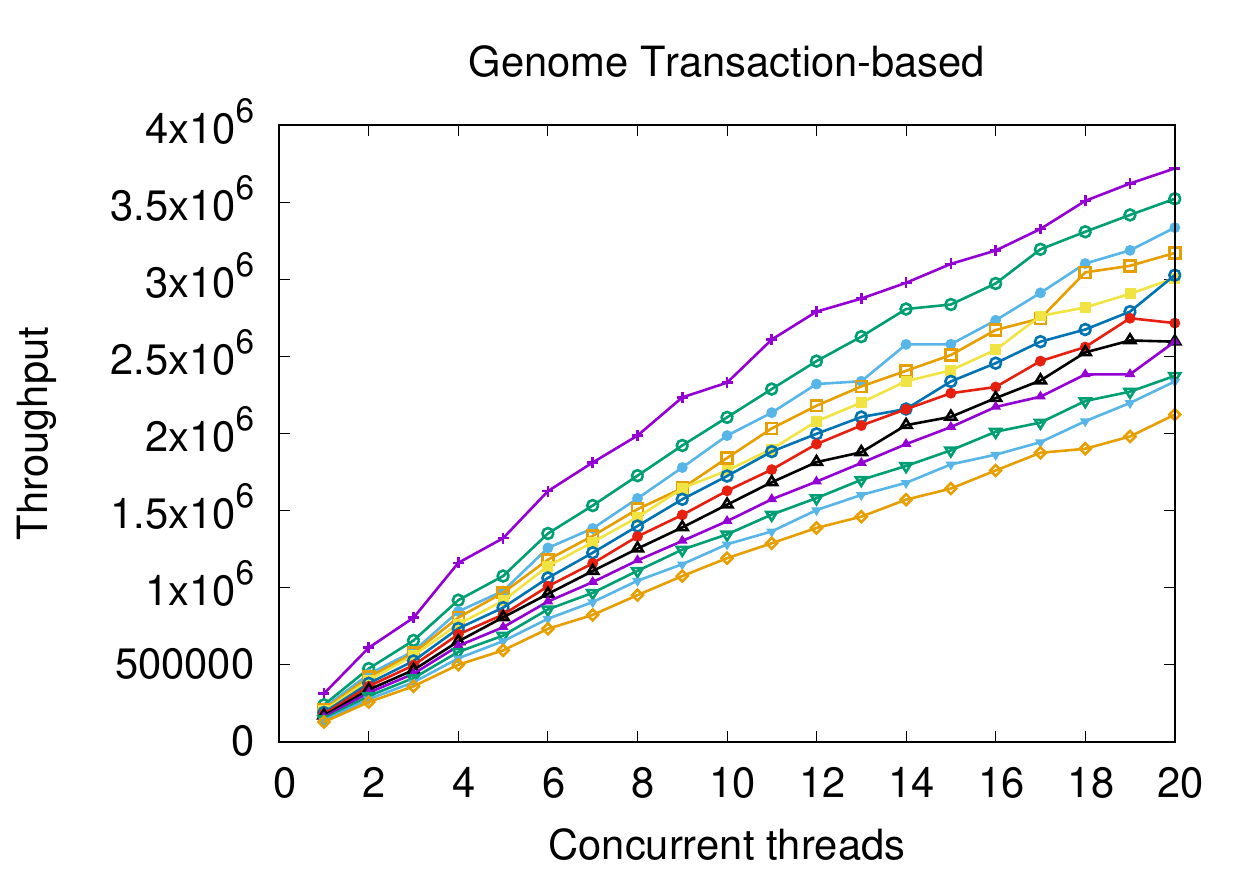}
\includegraphics[width=.32\textwidth]{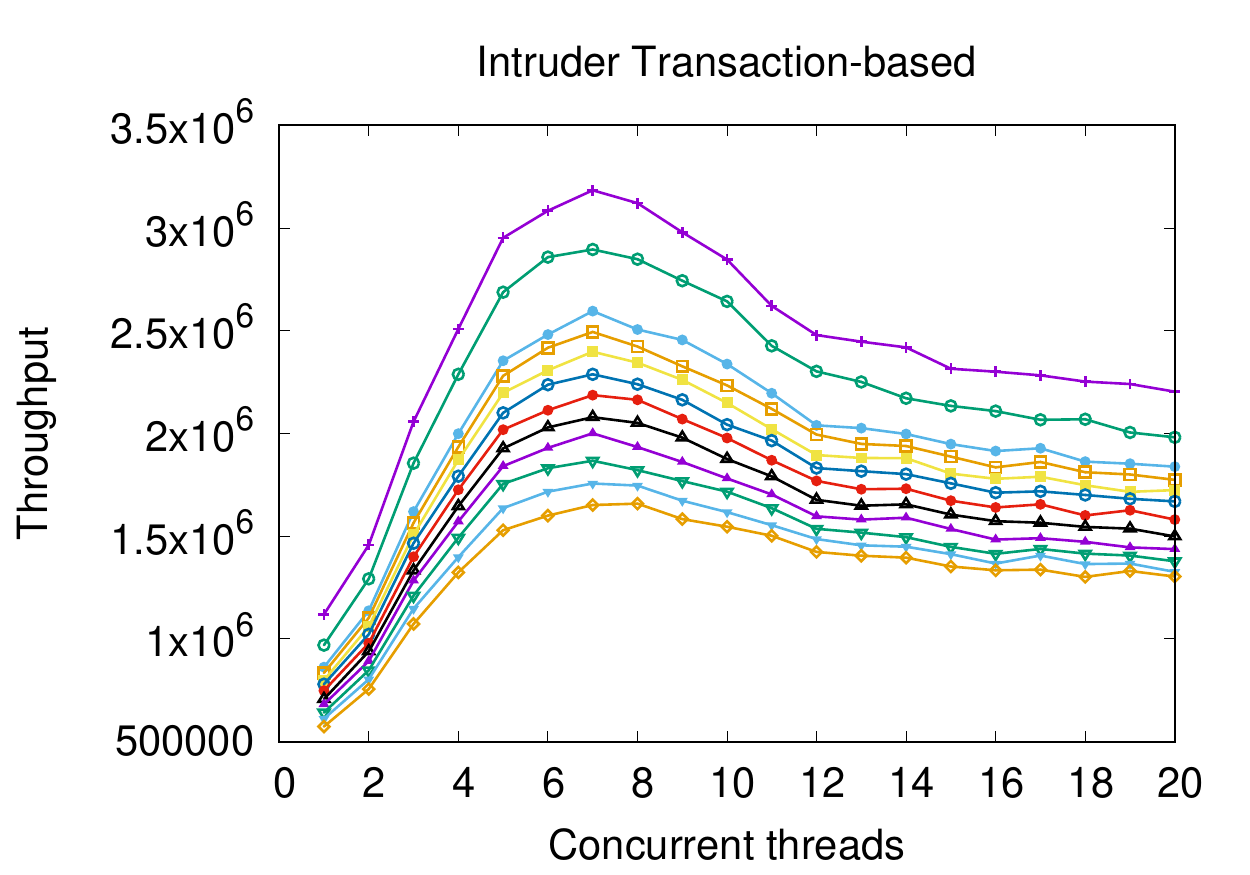}
\includegraphics[width=.32\textwidth]{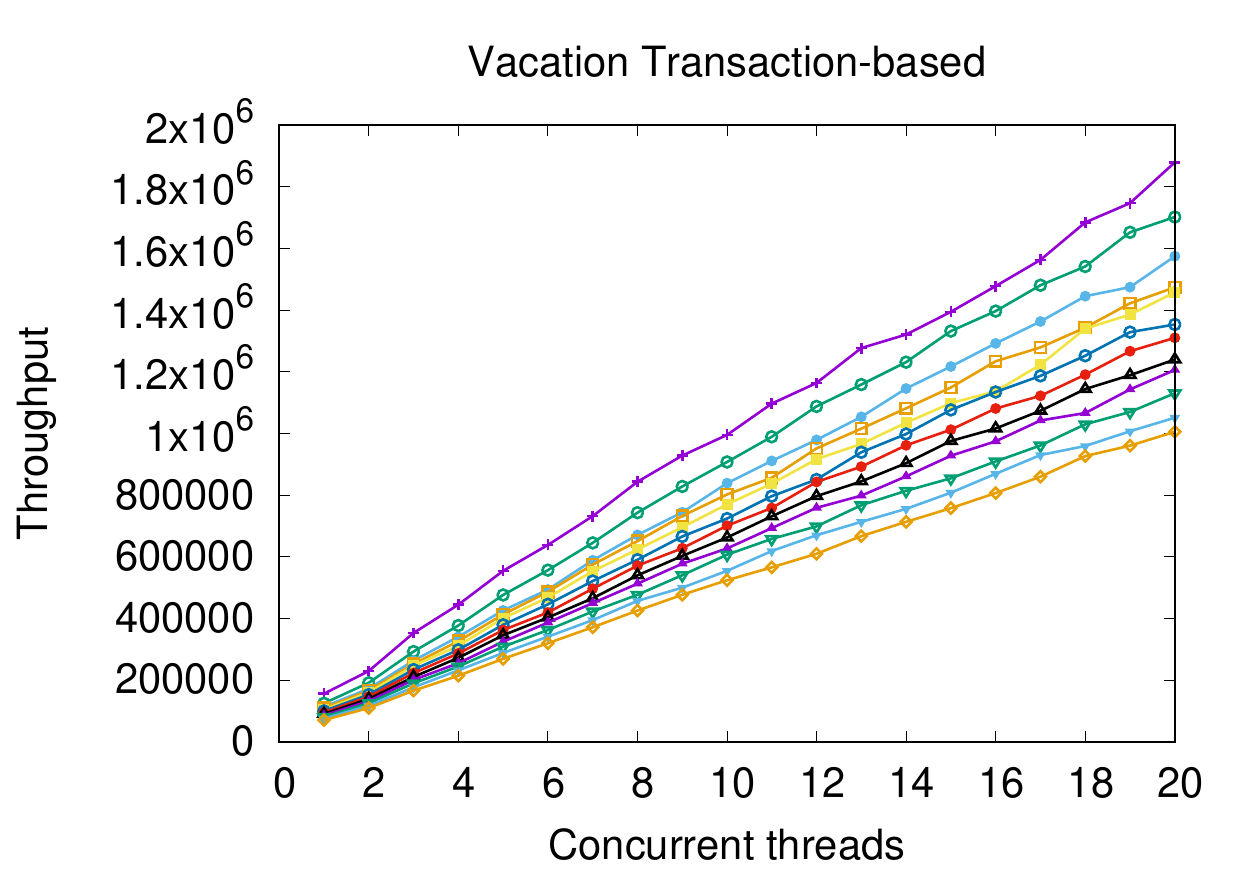}
\includegraphics[width=.32\textwidth]{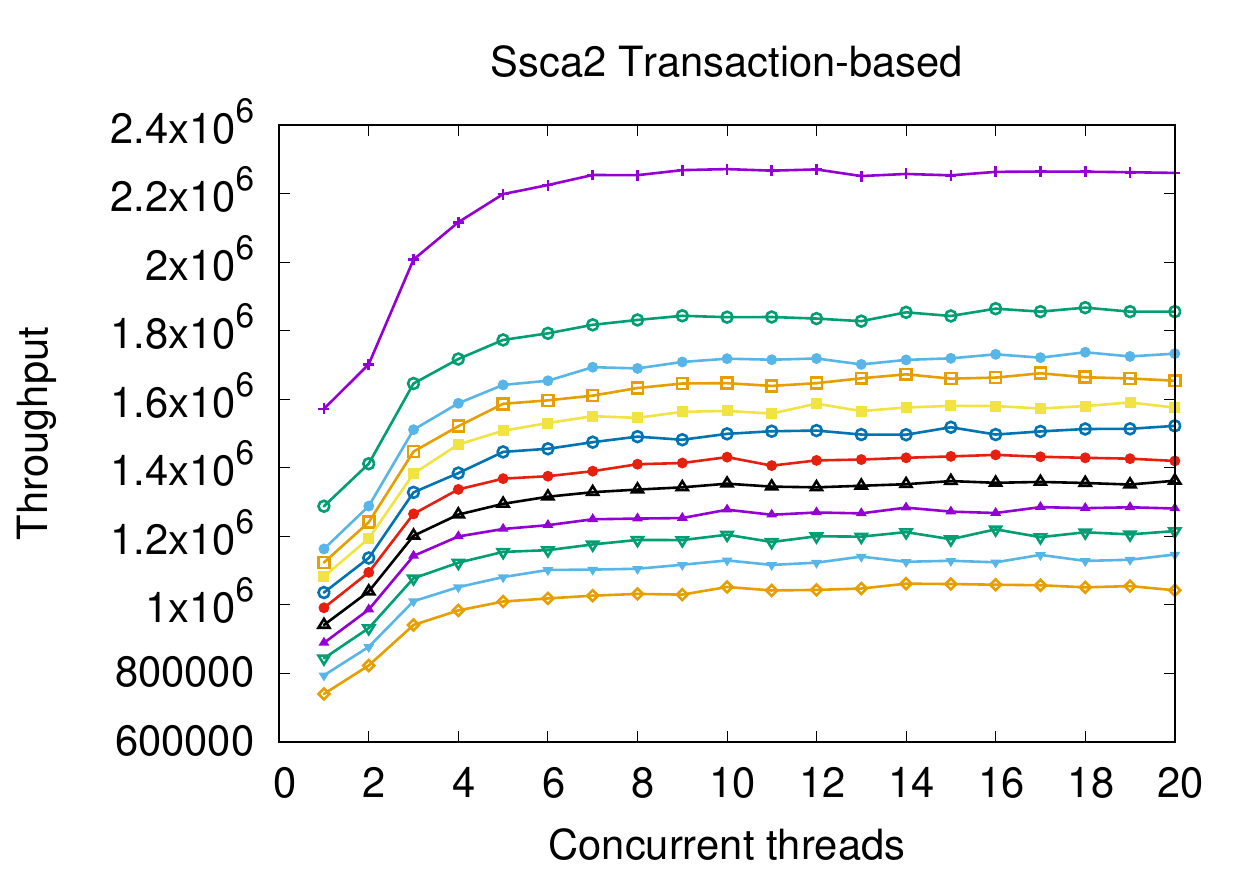}
\includegraphics[width=.32\textwidth]{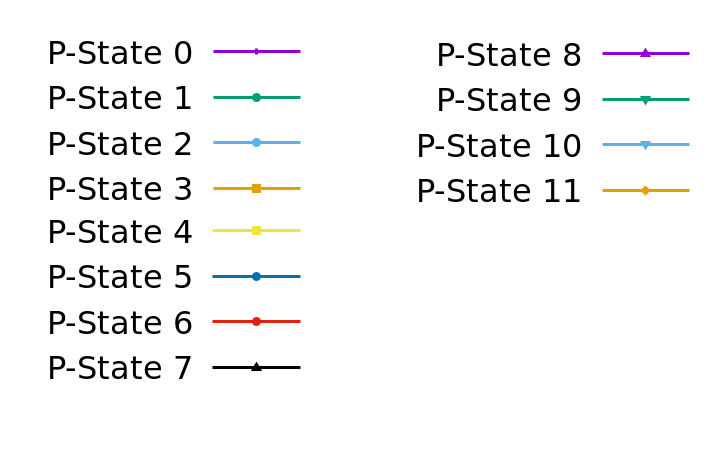}
\caption{Throughput vs. Number of Concurrent Threads}
\label{throughput_analysis}
\end{figure*}

In Figure \ref{throughput_analysis}, we report the results of an experimental study we conducted with three different multi-thread applications still taken from STAMP, namely Intruder, Genome, Vacation and Ssca2. We selected these applications since their scalability trends are very different. Also,
 in our experiments we considered two different implementations of the thread synchronization logic: a) a coarse-grained lock-based approach, where critical sections are synchronized by a single global lock, and b) a fine-grained approach based on software transactional memory, where shared data accesses are synchronized by transactions. We purposely used a coarse-grained locking scheme to evaluate our approach in various and antithetical scenarios, spanning from applications with very limited to very high scalability.
 

By the plots in Figure \ref{throughput_analysis}, the profile of the throughput curves confirm that there is an ascending part followed by an descending part. In some cases the ascending or the descending part may not exist. Also, the plots show that, when changing the application and/or the synchronization approach, the shapes of the throughput curves change. Particularly, the number of threads that provides the best throughput is generally different. Its range varies from 1 (in the case of workloads with very limited scalability, such as for Intruder Lock-based, Vacation Lock-based and Ssca2 Lock-based), up to 20 (in the case fully scalable workloads as Genome Transaction-based or Vacation Transaction-based). Notably, in some cases it is in the middle (as for Intruder Transaction-based, Genome Lock-based or Ssca2 Transaction-based). On the other hand, fixed the application and the synchronization approach, the throughput curves preserve the shape when varying \textit{P-state}. Curves appear proportionally translated, but the number of threads that provides the best throughput does not change, unless for small and unpredictable variations due to the measurement noise. Finally, the plots shows that, keeping fixed the number of threads, the throughput increases when decreasing \textit{P-state}.
We exploit these experimental findings to define the exploration-based technique presented in the next section.

\section{The Adaptive Power Capping Technique}
\label{technique}

The adaptive power capping technique we propose aims at finding the optimal configuration $(p, t)^*$, i.e. the configuration that provides the highest performance among the configurations in the set $\{(p, t)\}_{ac}$, assuming that it may change due to variations of the workload profile. The technique is based on an on-line tuning strategy that periodically performs an exploration procedure. The latter aims to identify the optimal configuration $(p, t)^*$ for the current workload profile, which is actuated until the exploration procedure restarts after a given period. During the exploration procedure, the power consumption and the throughput of the application are measured while moving along configurations within a given path, discarding the explored configurations that are not in the set $\{(p, t)\}_{ac}$. Then, the one with the highest throughput is selected. We should note that the number of threads $t^*$ of the optimal configuration may be different than the number of thread that provides the highest throughput for the specific application, since decreasing the CPU \textit{P-state} might provide an higher performance increase than increasing the number of threads. The procedure is able to identify the optimal configuration by exploring only a subset of configurations. In effect, we note that the full set of configurations may be very large, particularly when a large number of cores are available. Thus, reducing the exploration space is fundamental to implement an on-line exploration-based strategy. 

\subsection{The Exploration Procedure}
\label{subsection_exploration_procedure}

The exploration procedure takes as input a starting configuration $(p^s, t^s)$ and a power cap value $C$ and returns $(p, t)^*$. For the first execution of the procedure, the starting configuration is established by the user, while in next executions it corresponds to the output configuration of the previous one. We note that, based on the shapes of the throughput curves and the observations that we made in our preliminary analysis, a set of configurations can be excluded from the exploration, thus reducing the configuration exploration space. Specifically, if during the exploration:

\begin{enumerate}

\item \label{h0} a configuration  $(p^j, t^k)$ such that $thr(p^j, t^k) \leq thr(p^j, t^k-1)$ is found then all configurations $(p, t)$ where $t \geq t^k$, for whichever $p$, can be excluded (since we are in the descending part of the throughput curve and since the throughput curves preserve the shape while varying \textit{P-state}).
\item \label{h1} a configuration $(p^j, t^k)$ such that $pwr(p^j, t^k) \leq C$ is found then all configurations $(p, t^k)$ with $p > p^k$ can be excluded (since increasing \textit{P-state} reduces the application throughput).
\item \label{h2} a configuration  $(p^j, t^k)$ such that $pwr(p^j, t^k) > C$ is found then all configurations $(p, t)$ where $t \geq t^k$ and $p \leq p^k$ can be excluded (since decreasing \textit{P-state} or increasing the number of concurrent threads increments the power consumption).
\end{enumerate}
Based on the above observations, we built an exploration procedure divided in 3 phases, plus a final selection phase. The phases are described below. A graphical example is shown in Figure \ref{intruder_exploration}, which refers to an execution where the number of concurrent threads providing the highest throughput is equal to 15 and  C = 50 watts.

\noindent
The phases are the following ones:

\begin{figure}
\centering
\includegraphics[width=.5\textwidth]{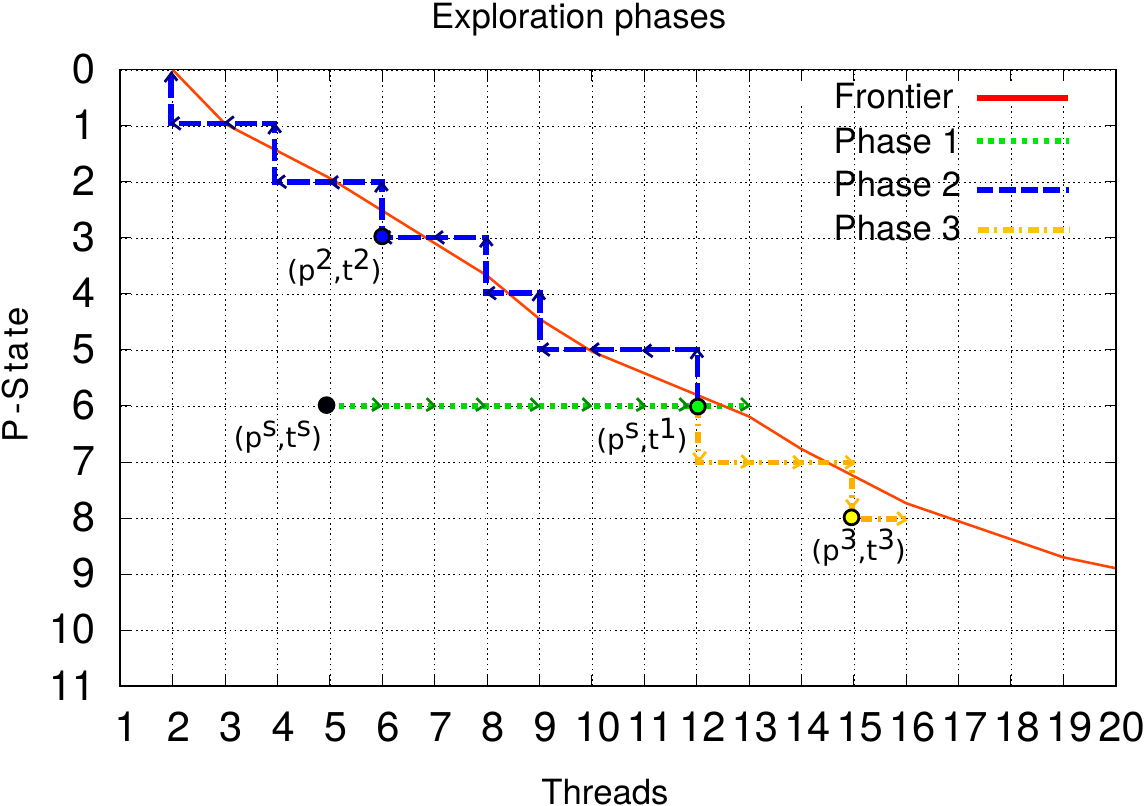}
\caption{Example of exploration phases performed by the basic strategy}
\label{intruder_exploration}
\end{figure}

\noindent
{\bf Phase 1:} this phase starts from the initial configuration $(p^s,t^s)$ and, keeping the \textit{P-state} fixed, aims at finding the number of threads providing the highest throughput without violating the power cap. We denote as $(p^s,t^1)$ the configuration returned by this phase. It performs a search inspired by the hill-climbing technique. Specifically, it increments by one the number of threads while the throughput increases and the power cap is not violated (since it is moving along the ascending part of the throughput curve), then it returns the configuration with the highest throughput within the power cap. If the throughput does not grow after the first increment or the power cap is violated, it starts decreasing the number of threads (since it is moving along the descending part of the throughput curve or the power consumption has to be reduced) until the throughput starts decreasing. Then, it returns the configuration with the highest throughput if it does not violate the power cap, otherwise if all the explored configurations violate the power cap or if the exploration reaches a number of threads equal to 1 it returns $(p^s,1)$. In the example in Figure \ref{intruder_exploration}, the exploration performed in Phase 1 is represented by the green line. It starts with $(p^s,t^s)=(6,5)$, then increases the number of threads and terminates when it explores configuration $(6,13)$ since it violates the power cap. It returns $(p^s,t^1)=(6,12)$, which is within the power cap.

\noindent
{\bf Phase 2:} This phase starts from the configuration returned by phase 1 $(p^s,t^1)$ and is executed only if this configuration does not violate the power cap (otherwise it jumps to the next phase). The goal of phase 2 is to continue the exploration for lower values of \textit{P-state} (we remark that lower values of \textit{P-state} lead to both higher core performance and higher power consumption). Specifically, it explores by moving from the current configuration $(p,t)$ to configuration $(p-1,t)$. If the latter configuration does not violate the power cap, it continues to reduce the value of \textit{P-state}. If the exploration reaches a configurations such that $pwr(p,t)>C$, it starts reducing the number of threads, thus moving to configuration $(p,t-1)$, then $(p,t-2)$ and so on (since decreasing the number of concurrent threads reduces the power consumption) until the power cap is not violated. After, it restarts the exploration by decreasing the value of \textit{P-state}. The exploration terminates when $p$ reaches $0$ and the current configuration does not violate the power cap, when it reaches configuration $(0,1)$, or when a configuration with $t=1$ violates the power cap. Then, the phase returns the explored configuration with the highest throughput within the power cap, that we denote as $(p^2,t^2)$, or configuration $(0,1)$. In Figure \ref{intruder_exploration}, the exploration of Phase 2 is shown by the blue line. It starts from $(p^s,t^1)=(6,12)$, then explores up to configuration $(0,1)$. It returns $(p^2,t^2)=(3,6)$.

\noindent
{\bf Phase 3:} This phase starts again from the configuration returned by Phase 1, i.e. $(p^s,t^1)$, and aims at continuing the exploration for higher values of \textit{P-state}. If the configuration returned by Phase 1 is such that $t^1$ is the number of threads providing the highest throughput and is within the power cap, Phase 3 is not executed (since decrementing the value of \textit{P-state} leads to lower throughput). Otherwise, it increments by one the value of \textit{P-state} and starts increasing the number of concurrent threads until the power cap is violated or the throughput decreases. In the former case, if the maximum value of \textit{P-state} has not been reached, it increments by one the value of \textit{P-state} and starts again incrementing the number of threads. In all the other cases the exploration terminates. Then, the phase returns the explored configuration, that we denote as $(p^3,t^3)$, with the highest throughput within the power cap, or configuration $(p_{max},t^1)$, where $p_{max}$ is the maximum value of \textit{P-state}. In Figure \ref{intruder_exploration}, the exploration of Phase 2 is represented by the yellow line. It starts from $(p^s,t^1)=(6,12)$, then explores up to configuration $(8,16)$, where it stops since the throughput decreases (in the example the number of concurrent threads providing the highest throughput is equal to 15). It returns $(p^3,t^3)=(8,15)$.

\noindent
{\bf Final phase:} this phase selects the configuration with the highest throughput between the configurations $(p^s,t^1)$, $(p^2,t^2)$ and $(p^3,t^3)$, which does not violate the power cap, or returns \textit{null} if none of them is within the power cap.

\subsection{Proof of Optimality}

In this subsection we prove that the proposed exploration procedure returns the optimal configuration, i.e. the configuration $(p,t)^*$ that provides the highest level of performance with a power consumption lower than the power cap. The proof assumes that the observations discussed in Section \ref{preliminary_study} always hold true. Specifically, we take as hypotheses that:
\begin{enumerate}
\item \label{hyp1} the shape of the throughput curve for each fixed \textit{P-state} and varying number of active threads is characterized by an initial ascending part, followed by a descending part. Also, one of these parts may be missing;
\item \label{hyp2} if $thr(p^j,t^k) > thr(p^j, t^k+1)$ then $thr(p,t^k) > thr(p, t^k+1)$ for each $p$ (the shape of the throughput curves preserve the shape while varying the \textit{P-state});
\item \label{hyp3} if $p^j<p^k$ then $thr(p^j,t)>thr(p^k,t)$ for each fixed $t$ (decreasing the \textit{P-state} with a fixed number of threads always increases the throughput);
\item \label{hyp4} $pwr(p,t) \geq pwr(p^j,t^k)$ for each $p<=p^j$ and $t>=t^k$ (decreasing \textit{P-state} or increasing the number of threads increases the power consumption);
\item \label{hyp5} the workload is static during the exploration procedure; 
\item \label{hyp6} the samples of throughput and power consumption obtained for each explored configuration are equivalent to their real values; 
\end{enumerate}
We should note that hypothesis \ref{hyp5} and \ref{hyp6} are necessary for any exploration-based solution that relies on data gathered at run-time. In particular, they guarantee that if the optimal configuration is explored it will also be selected by the algorithm as the best configuration.
Hypotheses \ref{hyp1}, \ref{hyp2}, \ref{hyp3} and \ref{hyp4}, as shown in Figure \ref{throughput_analysis}, reflect properties that appear to be valid for all the considered workloads. 
\begin{proof}
We can partition the search space defined by the configurations $(p,t)$ in three distinct sub-spaces, delimited by the starting configuration $(p,t)^s$, such that: 
\begin{itemize}
\item $p=p^s$;
\item $p<p^s$;
\item $p>p^s$;
\end{itemize}
We denote as the optimal configuration for the sub-space of configurations $S$, the configuration $(p,t)^q \in S$ that provides the highest performance while operating within the power cap compared to all the configurations $(p,t) \in S$. 
Considering that the sum of these three sub-spaces covers the complete space of configurations, the configuration that provides the highest performance between the optimal configuration of all the sub-spaces will be the optimal configuration $(p,t)^*$. Thus, proving that the exploration procedure finds the optimal configuration for each of these sub-space is equivalent to prove that it finds the optimal configuration for the whole space of configurations.

\noindent 
{$\mathbf{p=p^s:}$} Phase 1 of the exploration procedure starts from $(p,t)^{s}$ and searches for the number of active threads at \textit{P-state} $p^s$ that maximizes performance with power consumption within the power cap. Therefore, it explores the considered sub-space. Phase 1 is based on an hill-climbing optimization algorithm which generally finds the local optima, which might not be the best possible solution. However, for hypothesis \ref{hyp1}, the local optima is also the global optima as it is not possible for a non-global optima to exist for a function with a single ascending part followed by a single descending part or, in case one of those is missing, for a monotonic function. We should note that, unlike traditional hill-climbing algorithms, the selected configuration might not be the global optima as it might require a power consumption higher than the power cap. In this case, exploiting the shape of the throughput function, the configuration with the highest number of active thread with a power consumption lower than the power cap is selected which is clearly the optimal configuration for the sub-space as the optimum is always located either at the end of the ascending part or at 1 thread if the ascending part does not exist. Else, the global optima is selected. Thus, phase 1 selects the optimal configuration for the sub-space of configurations with $p=p^s$.

\noindent
{$\mathbf{p<p^s:}$} Assume that the optimal configuration $(p,t)^{k+1}$ for the sub-space of configurations with $p=k+1$ is known. The optimal configurations for the sub-space with $p=k$ must have $t^k<=t^{k+1}$ since if $t^{k+1}$ is the optimal number of threads for $p=k+1$ it must be that either: 
\begin{itemize}
\item throughput with $t=t^{k+1}+1$ is lower than with $t=t^{k+1}$ with $p=k+1$. Thus, for hypothesis \ref{hyp2}, all configurations $(p, t)$ where $t > t^{k+1}$, for whichever $p$, can not be optimal; 
\item $pwr(p^{k+1},t^{k+1}+1)>C$ which implies, for hypothesis \ref{hyp4}, that for $p=k<k+1$ all configurations with $t>=t^{k+1}$ would have a power consumption higher than the power cap and thus can not be optimal.
\end{itemize}
If $t^{k+1}=1$ we can already conclude that the optimum for the sub-space of configurations with $p=p^k$ is $(p=k, t=1)$. Differently, for $t>1$ we can state that the throughput at \textit{P-state} $k$ monotonically increases in the range from 1 to $t=t^{k+1}$. If $t^{k+1}>1$ it must be that $thr(p^{k+1},t^{k+1}) >=thr(p^{k+1},t^{k+1}-1)$ which implies for hypothesis \ref{hyp2} that $thr(p^{k},t^{k+1}) >=thr(p^{k},t^{k+1}-1)$ and consequently that the throughput curve with $p=k$ for $t<t^{k+1}$ is in the ascending part. 
Therefore, as performed by Phase 2, starting the exploration of the sub-space of configurations with $p=k$ from the configuration $p^{k},t^{k+1}$ and, if necessary, decreasing the number of threads until the power cap is reached assures that the the optimal configuration for the sub-space is explored. The configuration returned by phase 1---which is the optimal configuration for the sub-space with $p=p^s$---is used as base case. The sum of each sub-space of configurations with $\{p=j \mid j \in [0, p^s-1]$ is equal to the sub-space of configurations with $p<p^s$. Therefore, the configuration with the highest performance between the optimal configurations for each of this sub-spaces will be the optimal configuration for the entire sub-space with $p<p^s$.

\noindent
{$\mathbf{p>p^s:}$} Assume that the optimal configuration $(p,t)^{k-1}$ for the sub-space of configurations with $p=p^{k-1}=p^k-1$ is known. We can state that $thr(p^k,t) <= thr(p,t)^{k-1}$ for each $t<=t^{k-1}$ since:
\begin{itemize}
\item if $t^{k-1}$ is the optimal value of t for $p=p^{k-1}$, it must be true that $thr(p^k,t^{k-1})>=thr(p^k, t)$ for each $t<=t^{k-1}$ (hypotheses \ref{hyp2});
\item $thr(p^k,t^{k-1})<(p^{k-1},t^{k-1})$ (hypothesis \ref{hyp3}).
\end{itemize} 
In addition, we can state that if $thr(p^k,t^{k-1})>thr(p^k,t^{k-1}+1)$ then for each configuration $(p,t)$ with $p>p^k$ it must be true that $thr(p,t)<thr(p^k,t^{k-1})$ since:
\begin{itemize}
\item for hypothesis \ref{hyp2}, increasing the number of threads over $t^{k-1}$ does not improve the throughput for any \textit{P-state};
\item for hypothesis \ref{hyp3}, increasing the \textit{P-state} reduces the throughput.
\end{itemize}
Therefore, considering that $pwr(p^k,t^{k-1})<C$ (hypothesis \ref{hyp4}) and that $(p^k,t^{k-1})$ is included in the sub-space of configurations with $p<p^s$, if $thr(p^k,t^{k-1})>thr(p^k,t^{k-1}+1)$ then all configurations in the sub-space of configurations with $p<p^k$ cannot be optimal configurations of the sub-space with $p<p^s$.
Starting from the configuration returned by phase 1---which is the optimal configuration for $p=p^s$--- phase 3 decrements the \textit{P-state} and increments the number of threads until the power cap is violated or the throughput decreases, which assures that the optimal configuration for the sub-space is explored. If the throughput decreases when increasing the number of threads or when the maximum \textit{P-state} is reaches, phase 3 is completed. By induction, it explores the optimal configuration of each sub-space of configurations with $\{p=j \mid j \in [p^s+1, p^{max}]$, excluding the sub-spaces that we proved cannot contain the optimum. Therefore, the configuration with the highest performance between the optimal configurations of the considered sub-spaces will be the optimal configuration for the entire sub-space with $p>p^s$.
\end{proof}

\subsection{Time Complexity Analysis}
The time complexity of the exploration procedure is expressed as the number of exploration steps required by the procedure to return the optimal configuration $(p,t)^*$. Let $p^{tot}$ be the total number of \textit{P-states} supported by the system and $t^{tot}$ the maximum number of concurrent threads for the specific application, which, in HPC applications, is usually set equal to the number of physical/virtual cores available in the system. Considering that the exploration procedure does not explore the descending part of the throughput curve, we could also denote $t^{tot}$ as the maximum number of concurrent threads that provide, for at least a portion of the execution time, the highest performance for the specific application run on the specific hardware. We analyze the time complexity of each exploration phase separately:
\begin{itemize}
\item \textbf{phase 1}: each configuration with a different number of concurrent threads and $p=p^s$ is explored at most once, thus the time complexity is $\mathcal{O}(t^{tot})$;
\item \textbf{phase 2}: starting from a configuration $(p,t)$, phase 2 either reduces the value of $p$ or reduces $t$. Starting from the configuration returned by phase 1, it can reduce $p$  at most $p^{tot}$ times, and reduce $t$ at most $t^{tot}$ times. Thus, the time complexity of phase 2 is $\mathcal{O}(p^{tot}+t^{tot})$;
\item \textbf{phase 3}: starting from a configuration $(p,t)$, phase 3 either increments the value of $p$ or increments $t$. Thus, for the same reasoning used in phase 2, the time complexity of phase 3 is $\mathcal{O}(p^{tot}+t^{tot})$;
\end{itemize} 
Therefore, the overall time complexity of the exploration procedure is  $\mathcal{O}(p^{tot}+t^{tot})$.

\subsection{The Enhanced Tuning Strategy}
\label{enanched_strategy}

In this section we present an enhancement of the tuning strategy that allows to further improve performance and reduce the power cap violation probability. It profits by the possible gap between the power cap value and the power consumption of configuration $(p, t)^*$ which is due to the discrete domain of power consumption values of the different configurations. Specifically, it is unlikely that $pwr(p, t)^*$ is exactly equal to $C$. Rather we can have $C-pwr(p, t)^*>0$. Statistically, the greater the difference of power consumption between adjacent configurations, the larger $C-pwr(p, t)^*$. To reduce the performance penalization due to this gap, the enhanced tuning strategy relies on continue fluctuations between two configurations (rather that remaining always in $(p, t)^*$) along the time interval between the end of the exploration procedure and the start of the next one. All the phases are equal to the previous tuning strategy, except that an additional configuration $(p, t)^H$ is selected. $(p, t)^H$ is the configuration with higher throughput than $(p, t)^*$ (if any) such that the ratio between throughput and power consumption is the largest one among the explored ones. Thus, it is the configuration with the highest efficiency in terms of throughput over power consumption. We note that,  since $(p, t)^*$ is the configuration within the power cap with the highest throughput, then $(p, t)^H$ is a configuration that violates the power cap.

At the end of the exploration procedure, the enhanced strategy continuously fluctuates between $(p, t)^*$  and $(p, t)^H$ in order to take advantage of the higher throughput of configuration $(p, t)^H$, but avoids that the average power consumption, over a given time window $w$, overcomes $C$. To this aim, if the average power consumption overcomes $C$, then configuration $(p, t)^*$ is set. Conversely, when the average power consumption falls below $C$, configuration $(p, t)^H$ is set, and so on. To limit the fluctuation frequency, an upper and a lower tolerance threshold, $C+l$ and $C-l$ is used. In real scenarios, the length $w$ can be set equal to the actual time window used to calculate the power consumption of the machine.

Another factor that may impact the effectiveness of our technique is the variation over time of the power consumption of the selected configurations. For example, $pwr(p, t)^*$ may change due to variations of the workload profile, thus leading to power cap violations. If this happens, with our tuning strategy it may not be detected until the next exploration procedure starts. To limit the effect of this delay on the power cap violation, the enhanced tuning strategy selects a third configuration, that we denote as $(p, t)^L$. It is the configuration with lower power consumption than $(p, t)^*$ (if any) with the highest efficiency in terms of throughput over power consumption. Thus, if $pwr(p, t)^*$ overcomes $C$, then the strategy fluctuates between $pwr(p, t)*$ and $pwr(p, t)^L$ rather than between $(p, t)^*$ and $(p, t)^H$. This allows to reduce the probability of power cap violation until the workload profile variation is such that $pwr(p, t)^L<C$. Similarly, for the same goal of promptly adapting to workload variations, if $pwr(p, t)^L>C$ ($pwr(p, t)^H<C$), the \textit{P-state} of all configurations is shifted up (down) by one.

\section{Experimental Results}
\label{experimental}

In this section, we presents the results of an experimental study we conducted to asses the proposed power capping technique. As in previous studies on power capping (e.g. \cite{Reda:2012,Lefurgy:2008}), we consider two evaluation metrics, the application performance and the average power cap error. The latter is the average difference between the power consumption and power cap value along time intervals where the power cap is violated. We run experiments for all application scenarios that we considered in our preliminary study (see Section \ref{preliminary_study}). Thus we use Intruder, Genome,Vacation and Ssca2 as benchmark applications from STAMP, with both locks and transactions as the synchronization method. These applications were specifically selected to cover a wide range of different scalability scenarios.
We compared our technique with:
\begin{enumerate}
\item a reference power capping technique, referred to as baseline, that selects the configuration with the lowest \textit{P-state} from the set of configurations with the highest number of threads among the configurations with power consumption lower than the power cap. It implements the selection strategy proposed in \cite{Reda:2012};
\item a technique, referred to as dual-phase, that initially tunes the number of threads starting from the lowest \textit{P-state}, and subsequently tunes the CPU \textit{P-state} keeping the number of threads fixed. The initial phase is equivalent to phase 1 of the proposed exploration procedure. The selection strategy of this technique is similar to the one presented in \cite{Zhang_2016}.
\end{enumerate}
The comparison with the first technique allows to quantify the performance benefits achievable by properly allocating the power budget taking into consideration the scalability of the specific multi-threaded application. Additionally, we considered the dual-phase technique in the evaluation to quantify the possible performance benefits achievable by exploring the whole bi-dimensional space of configurations over two distinct mono-dimensional explorations, which might not find the optimal configuration. We should note that, despite exploring a larger set of configurations, the proposed technique has the same time complexity of the dual-phase technique.  


\begin{figure*}[h]
\centering

\includegraphics[width=.32\textwidth]{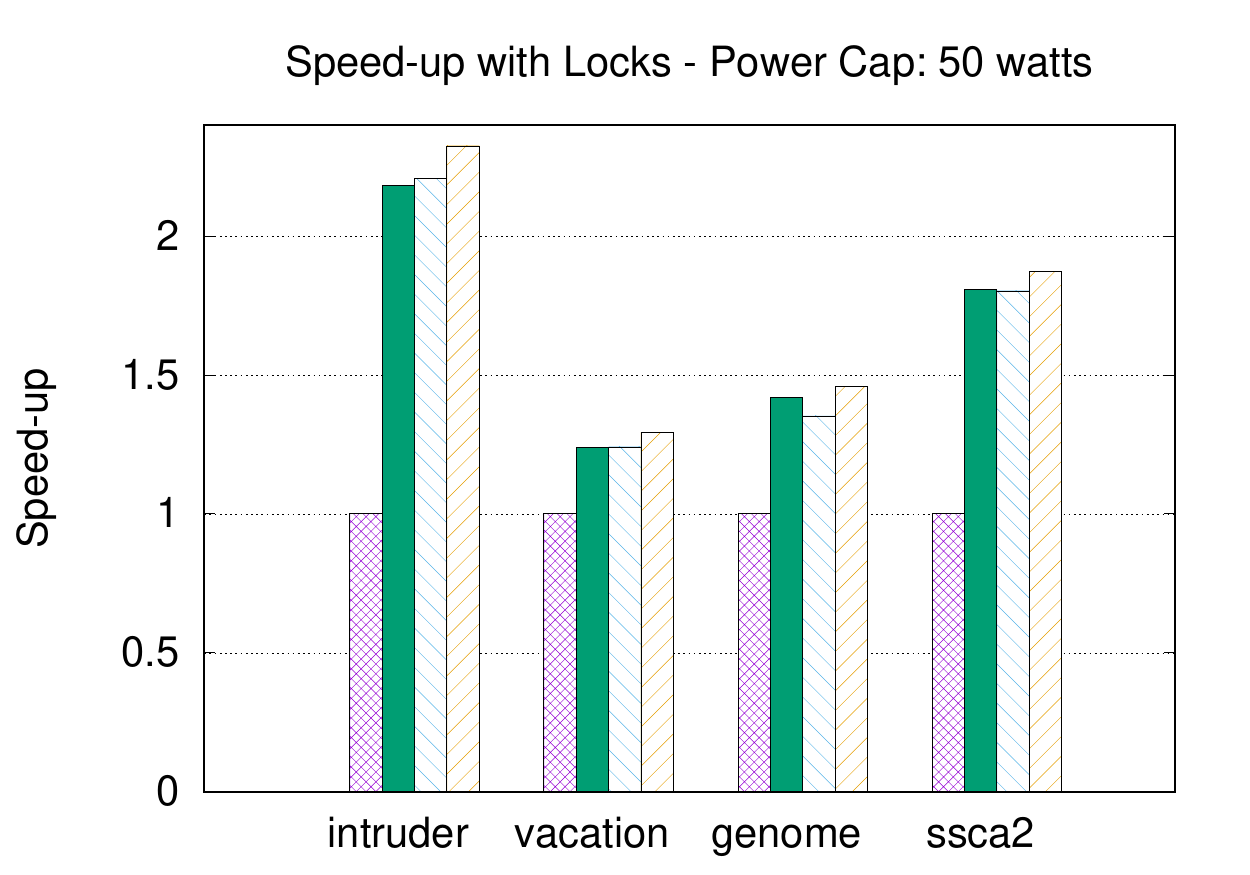}
\includegraphics[width=.32\textwidth]{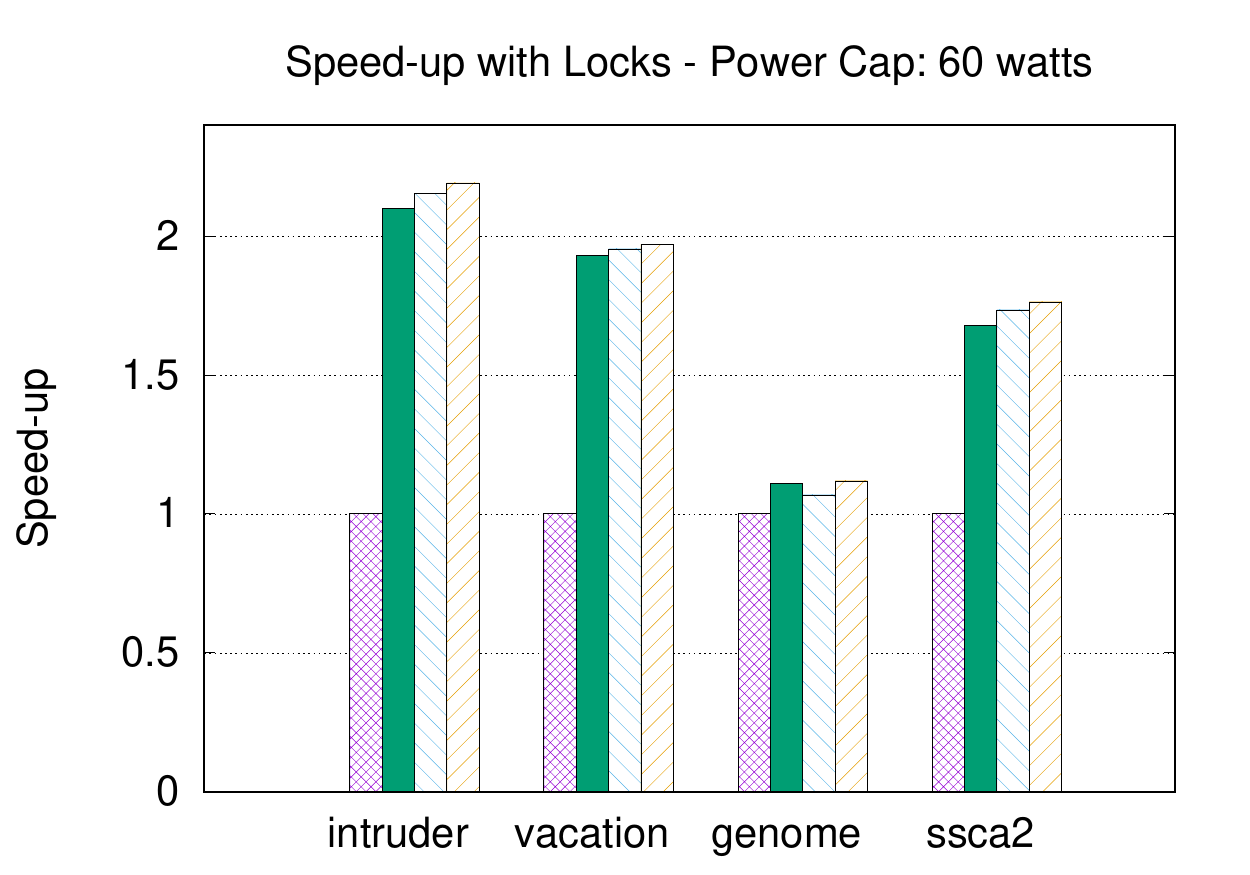}
\includegraphics[width=.32\textwidth]{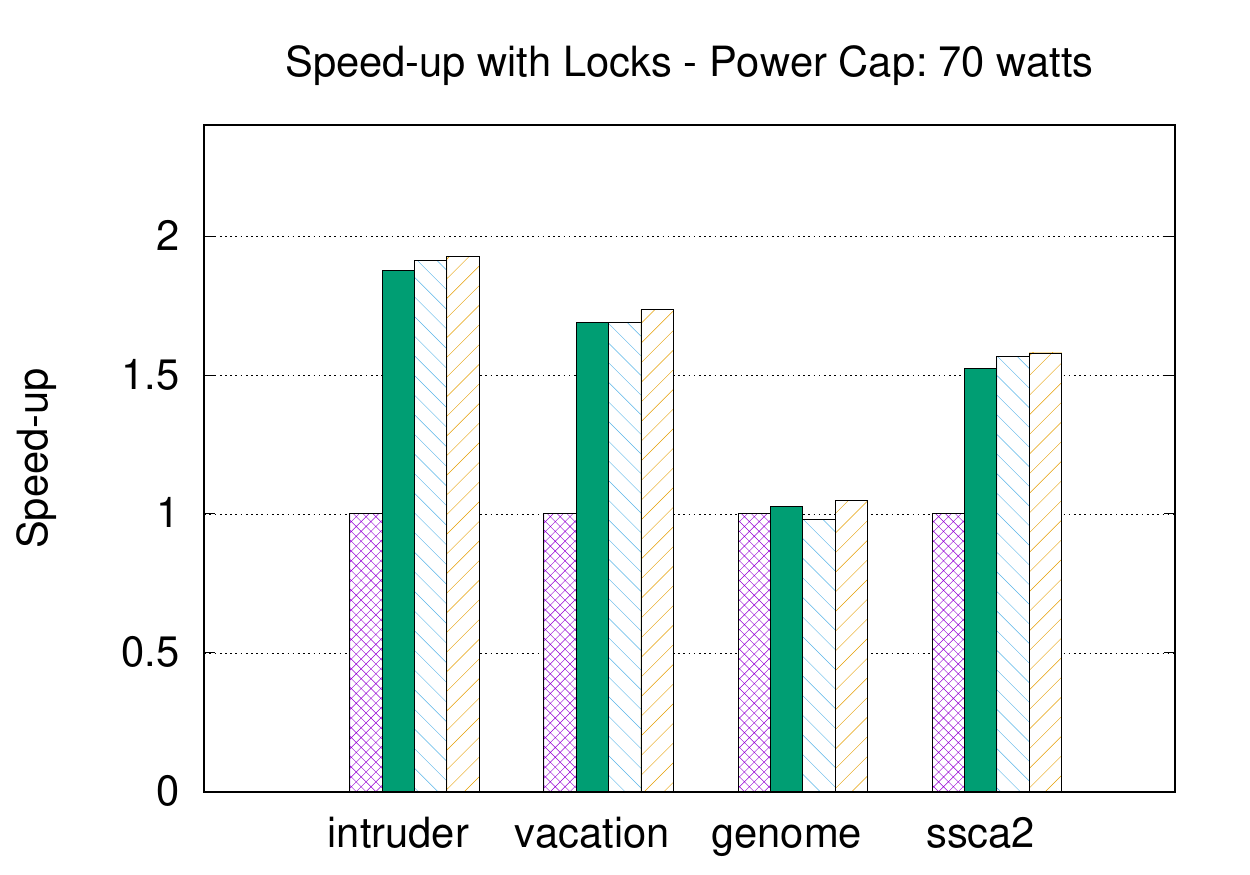}

\includegraphics[width=.32\textwidth]{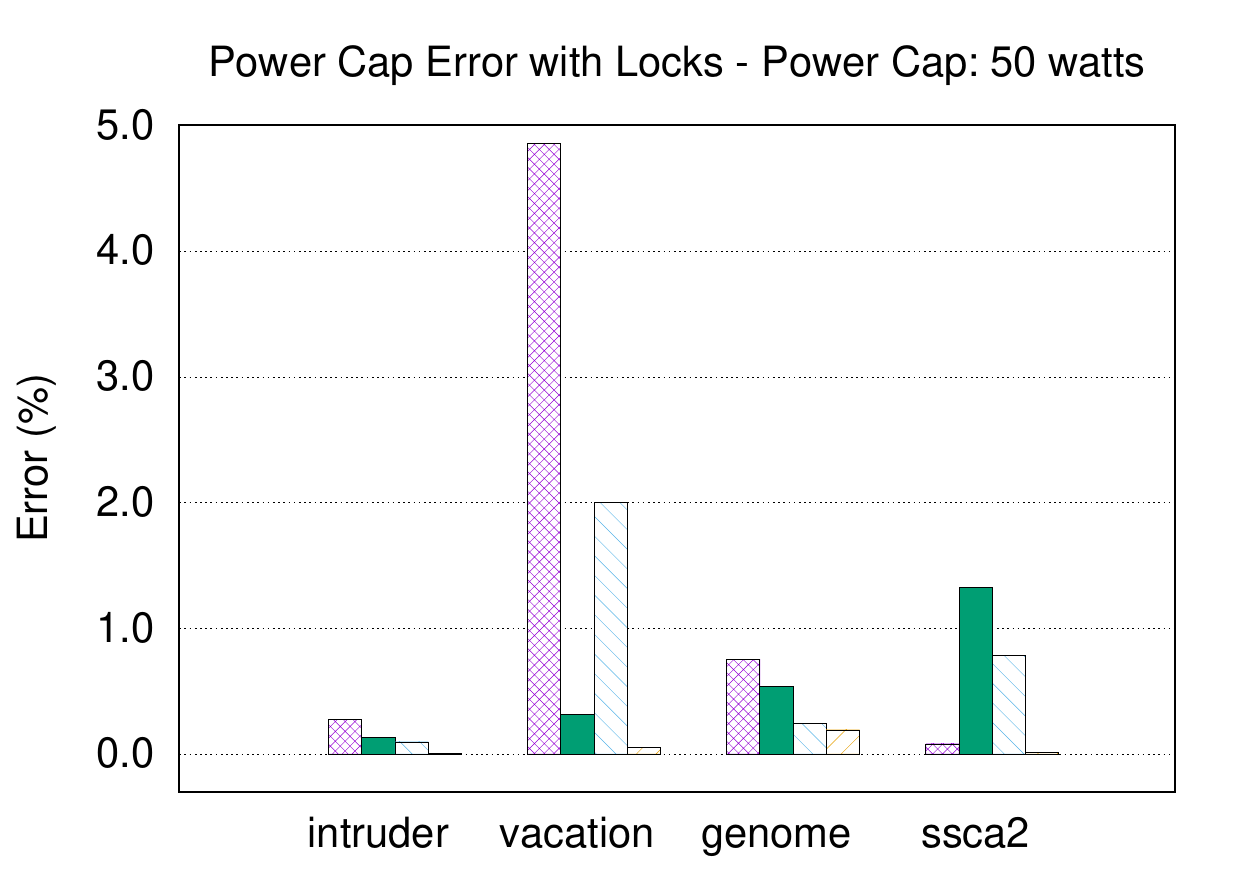}
\includegraphics[width=.32\textwidth]{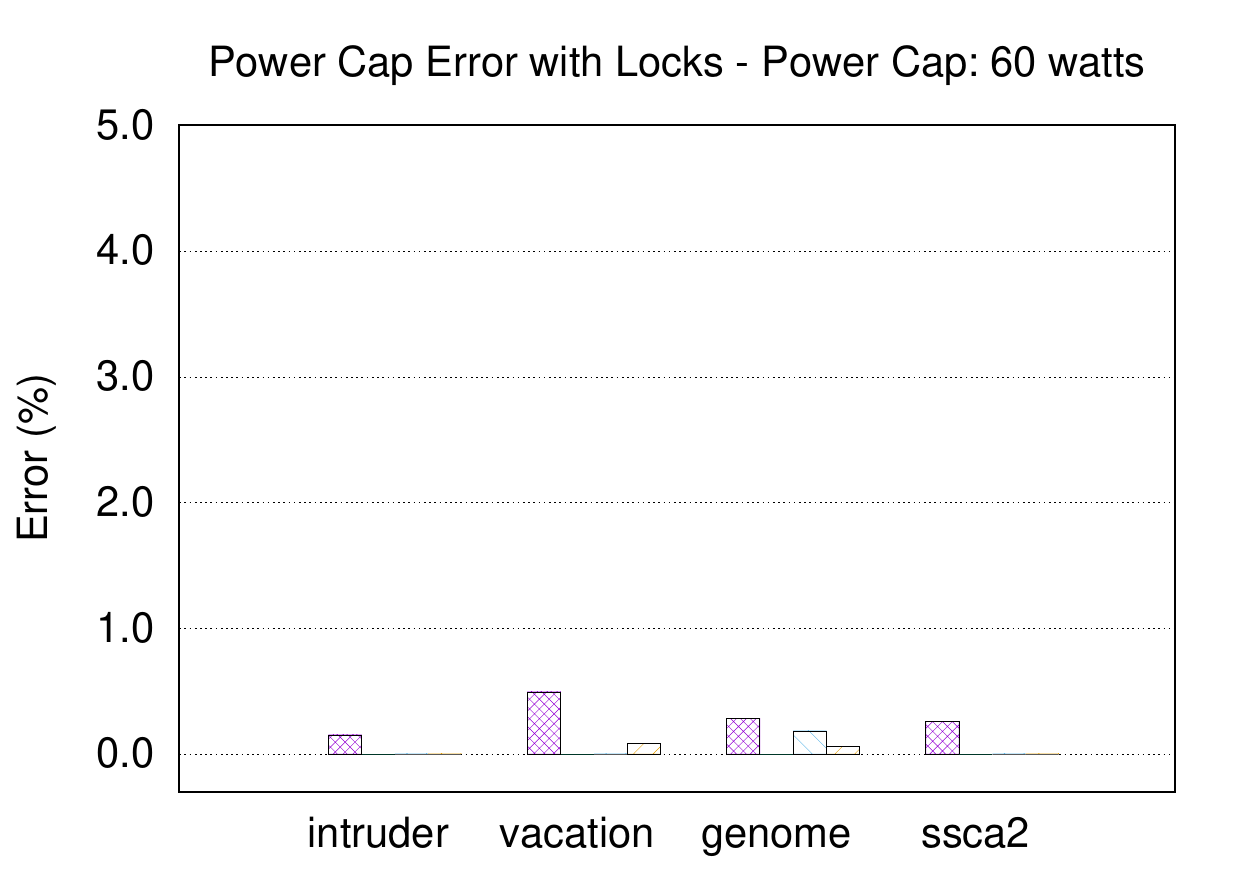}
\includegraphics[width=.32\textwidth]{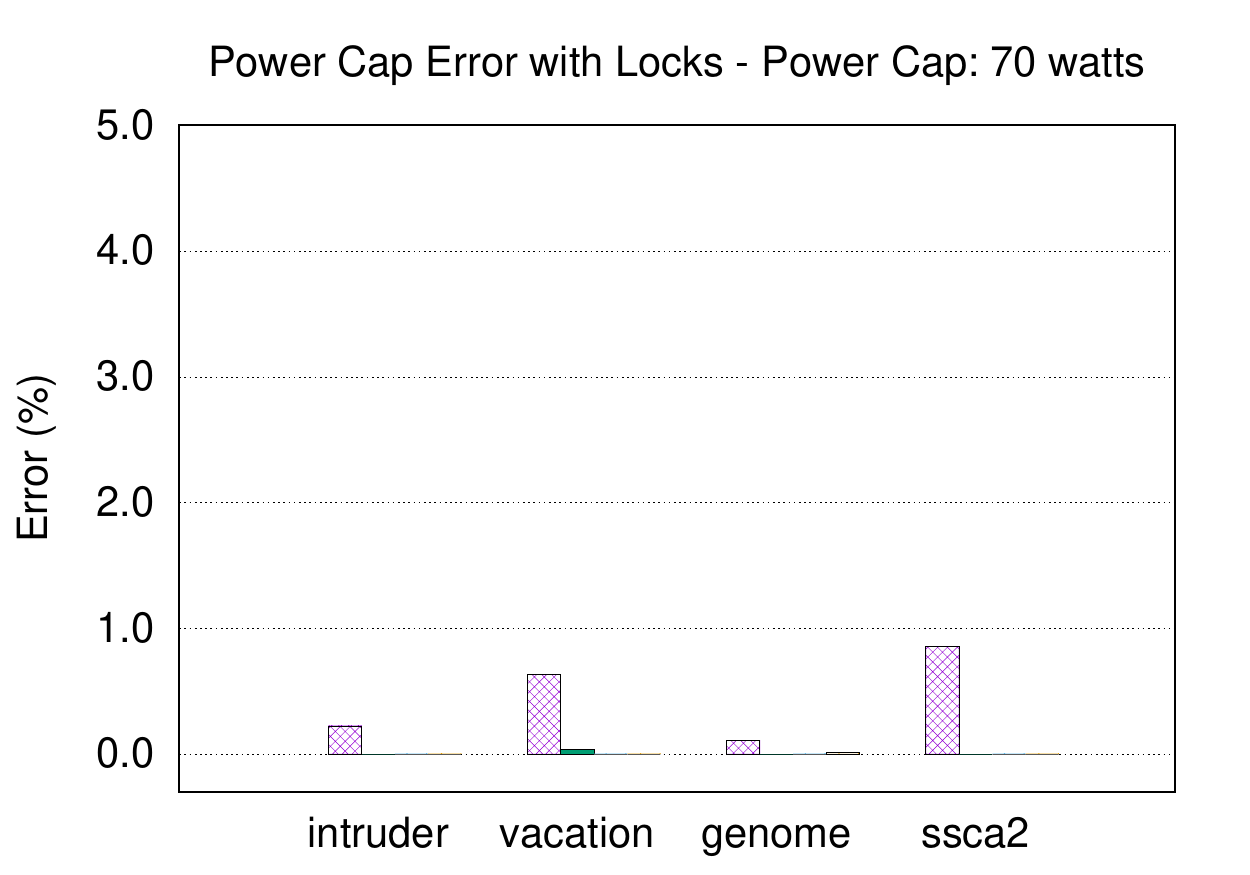}

\includegraphics[width=.8\textwidth]{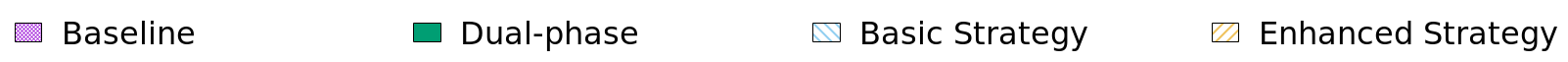}

\caption{Throughput Speed-up and Power Cap Error with Locks}
\label{results_with_locks}
\end{figure*}

\subsection{Implementation details}

We developed a controller module that implements our technique and the baseline technique.\footnote{See {\tt github.com/StefanoConoci/STMEnergyOptimization}}
All software of our experimental study, including benchmark applications, is developed in C language for Linux.
The controller module alters the number of concurrent threads exploiting the \emph{pause()} system call and thread-specific signal for reactivation. The CPU \textit{P-state} is regulated through the \emph{cpufreq} linux sub-system, while energy readings are obtained from the \emph{powercap} sub-system. Both these sub-systems are included by default in recent versions of the linux kernel and expose their respective interface through the \emph{sys} virtual file system.

The exploration procedure relies on statistical results of the previous step, such as average power consumption and throughput, to define the following configuration to explore. Each step of statistics collection is determined by a fixed amount of units of work processed. We cannot rely on application independent metrics, such as the number of CPU retired operation, since it would also consider instructions related to spin-locking or aborted transactions that do not provide execution progress. For applications based on locks we defined the unit of work as the execution of one critical section guarded by a global lock. Differently, for transactions we define the unit of work as one commit.
The statistics are collected in a round-robin fashion by all the active threads to reduce execution overhead and provide NUMA-aware results in modern multi-package systems.

For the executions presented in the experimental results, we set the units of work per step to 5000, resulting in tens of milliseconds per step for all the considered applications and synchronization method. In addition, we set to 150 the number of steps required to restart the exploration procedure after the conclusion of the previous.

\begin{figure*}[h]
\centering

\includegraphics[width=.32\textwidth]{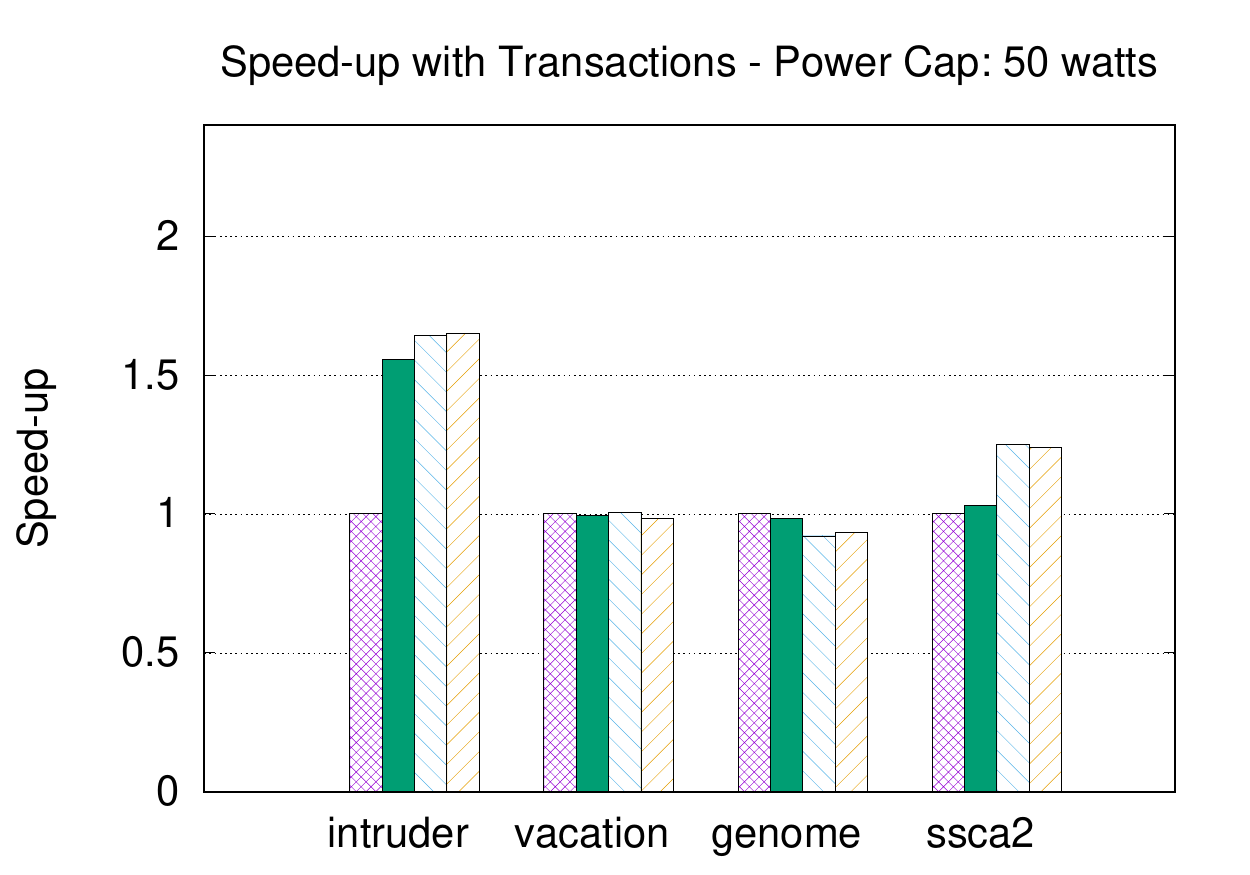}
\includegraphics[width=.32\textwidth]{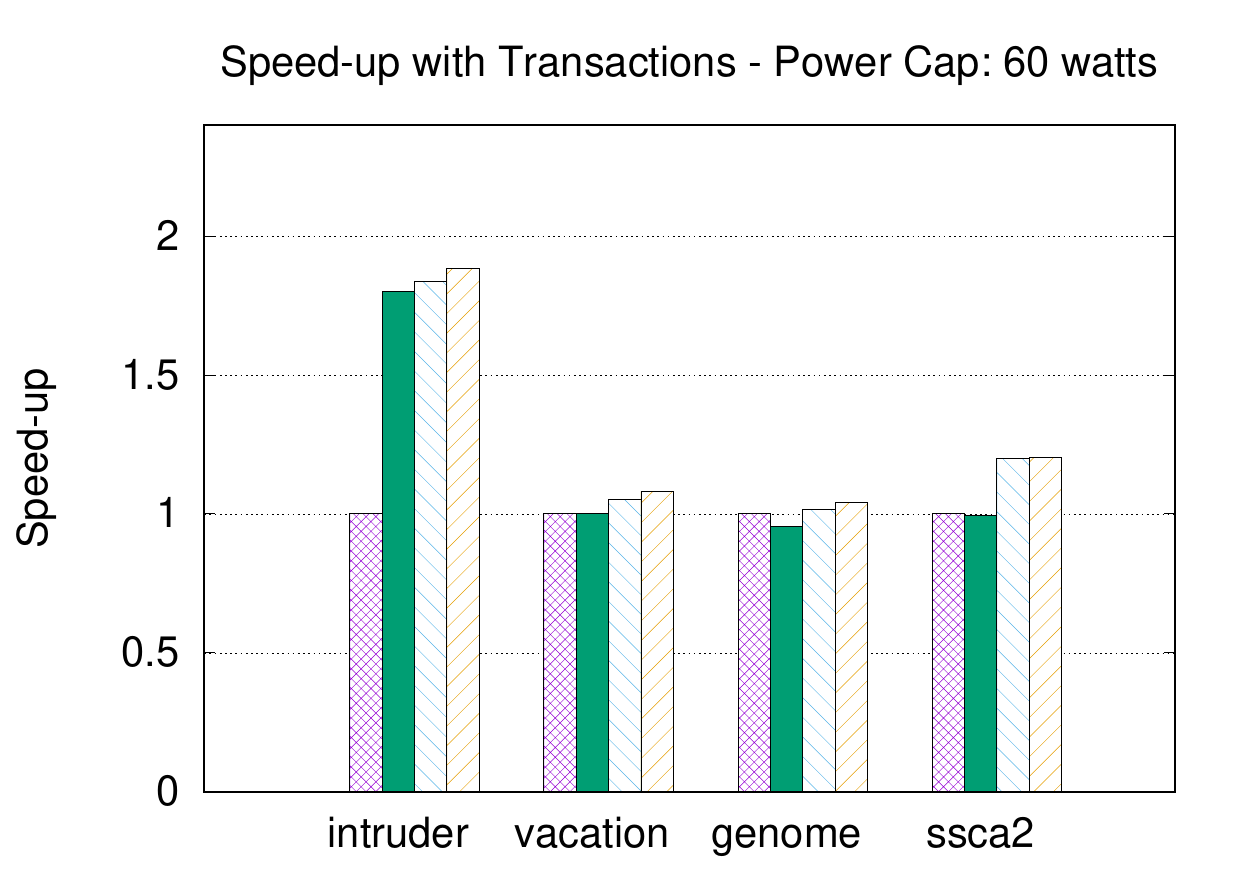}
\includegraphics[width=.32\textwidth]{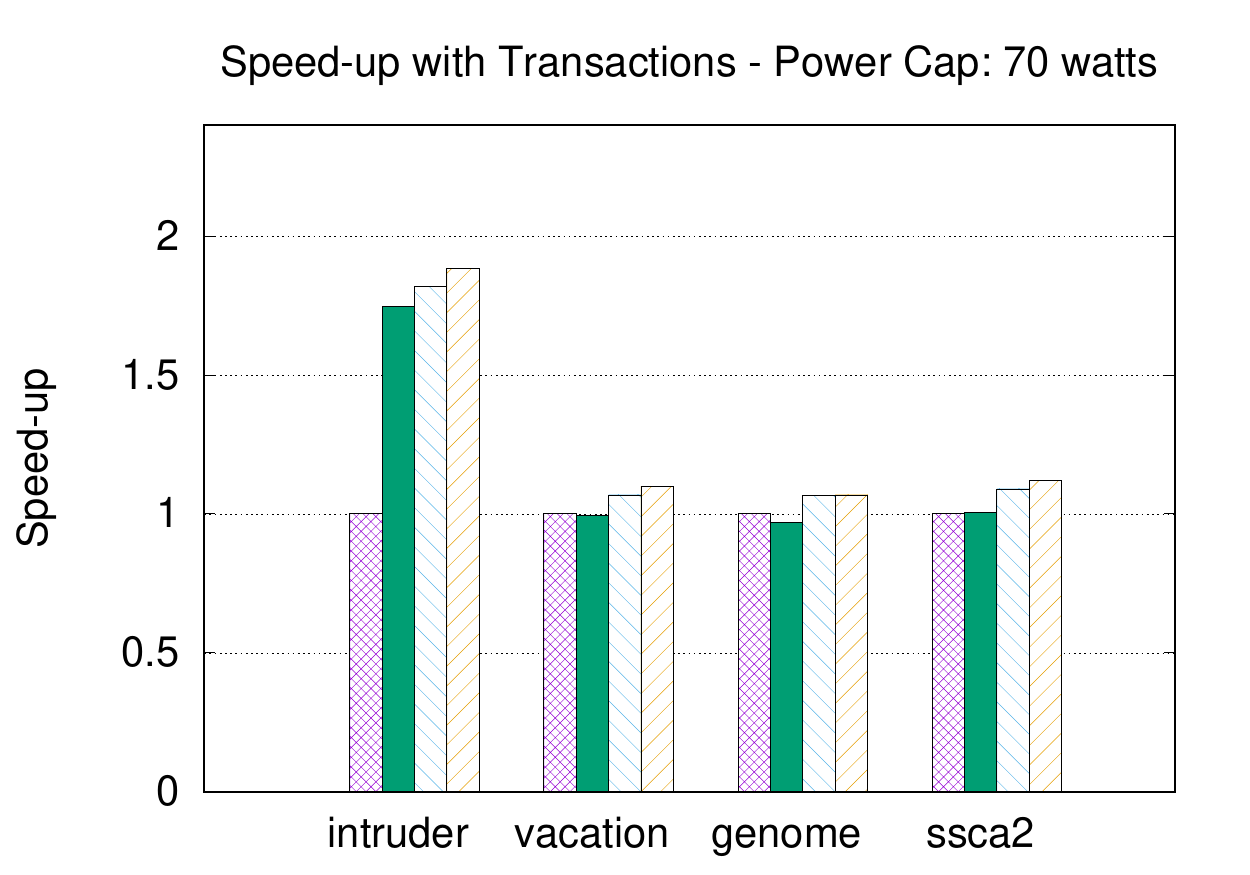}

\includegraphics[width=.32\textwidth]{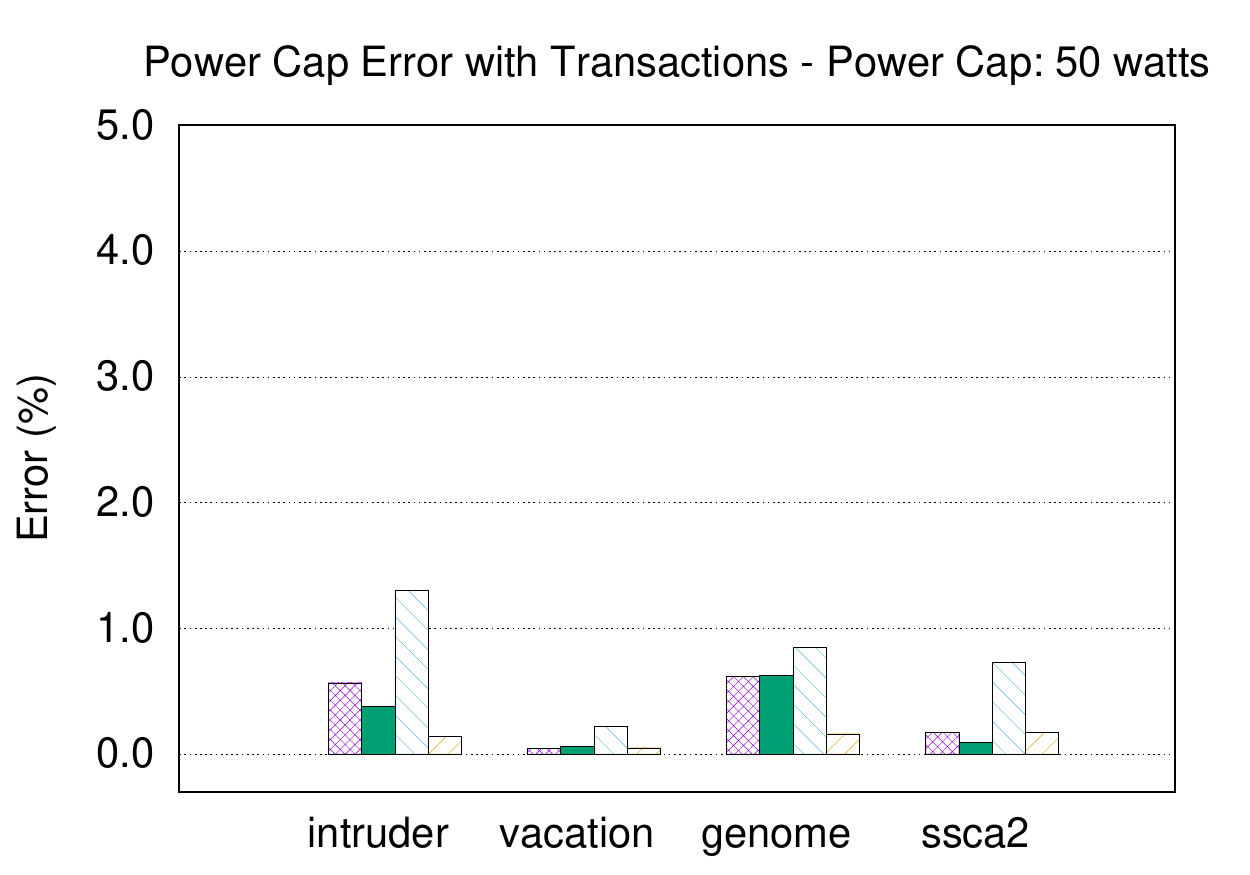}
\includegraphics[width=.32\textwidth]{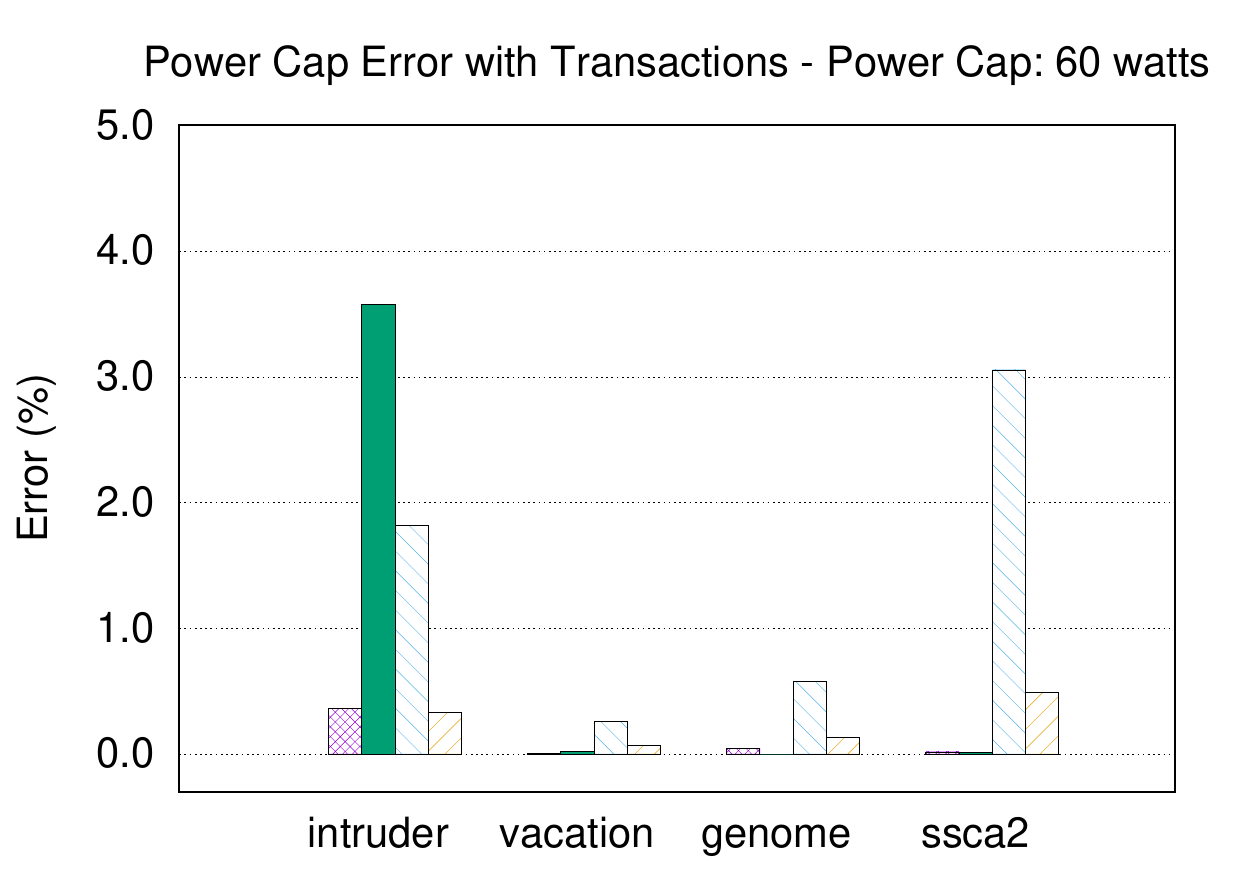}
\includegraphics[width=.32\textwidth]{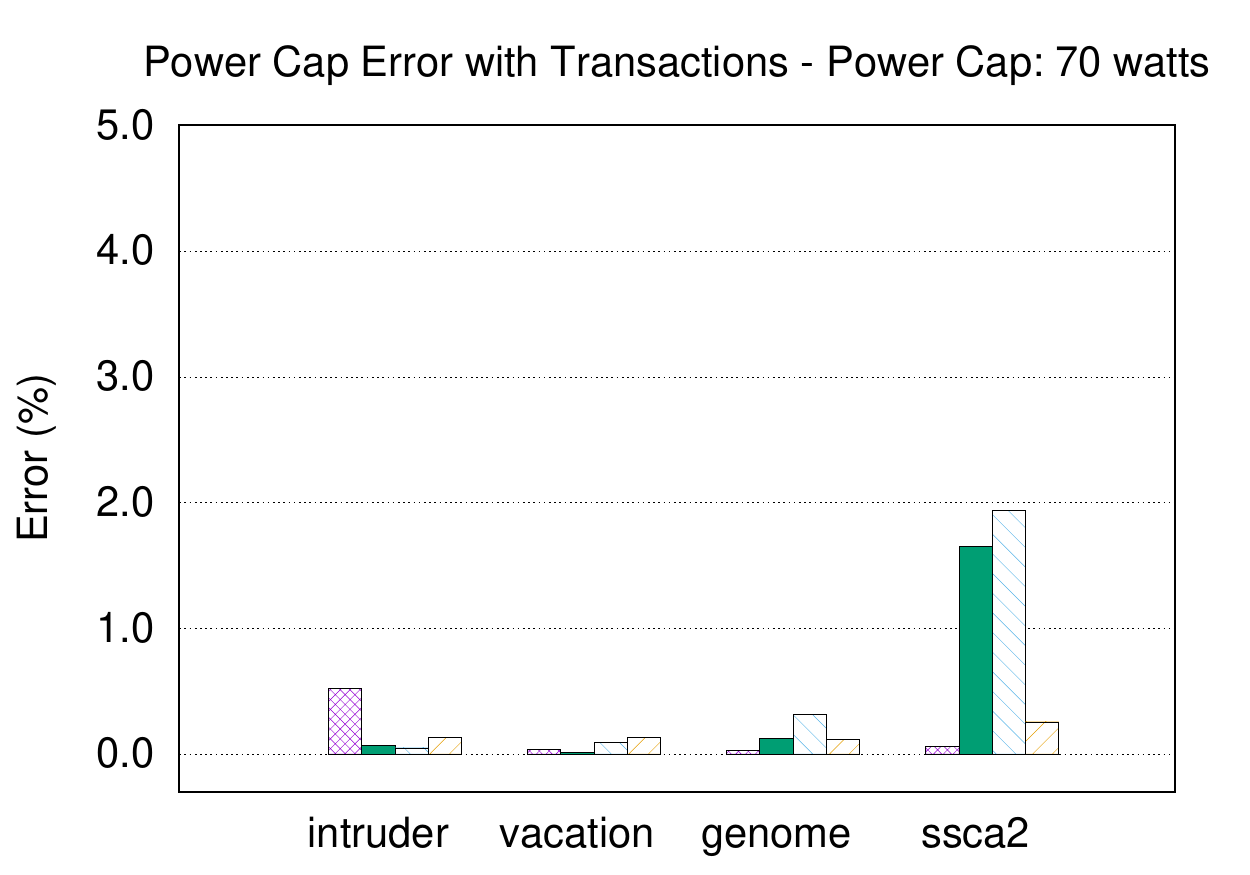}

\includegraphics[width=.8\textwidth]{key.png}

\caption{Throughput Speed-up and Power Cap Error with Transactions }
\label{results_with_stm}
\end{figure*}

\subsection{Experimental results}

We consider both the tuning strategies of our technique referred as basic strategy and enhanced strategy. We analyze the performance results of our strategies in terms of speed-up with respect to the throughput of the baseline technique. As anticipated, we also compare the average power cap error. For each test case, we present the results with three different power cap values, i.e. 50, 60 and 70 watts.

Results for the case of lock-based synchronization are reported in Figure \ref{results_with_locks}. Overall, the results show an evident performance improvement with both strategies of our technique with respect to the baseline technique. Only for the case of Genome the performance is comparable. In the best cases, i.e. with Intruder, the performance improvement reaches 2.2x (2.32x) and 2.15x (2.19x) for the basic (enhanced) strategy when the power cap is equal to 50 and 60 watts respectively, and it is close to 1.9x for both the proposed strategies with power cap set to 70 watts. The enhanced strategy further improves performance compared to the baseline technique by up to 12.5\% in Intruder at 50 watts, and by 5.3\% on average. For lock-based synchronization, the results of the dual-phase technique are similar to those achieved by the basic strategy.

As for the power cap error, with both the strategies of our technique and the dual-phase technique, it is clearly reduced compared to the baseline. Also, the results show that with the enhanced strategy in many cases there is a reduction of the power cap error compared to the basic strategy. Indeed, except for the case of Vacation with power cap equals to 60 watts, where it is increased by less than 0.1\%, the error with the enhanced strategy is lower. In the best case it is about 0.1\%, while it is about 2\% and 4.8\% with the basic strategy and the baseline technique, respectively.

Results for the case of transaction-based synchronization are reported in Figure \ref{results_with_stm}. Overall, the performance results confirm the advantage of our technique compared to the baseline technique. However, with transactions the speed-up is generally slightly lower than with locks. In the best cases, it reaches about 1.9x. Also, there is one case (with Genome and power cap = 50 watts) where it is slightly less that 1 with both the strategies. As for the power cap error, it increases with the basic strategy compared to the case with locks, overcoming the error of the baseline technique in most of the cases. However, it does not overcome 2\% in all cases. The error is considerably reduced with the enhanced strategy. Particularly, it is clearly lower than the baseline technique with all applications when the power cap is equal to 50 watts and with Intruder when the power cap is equals to 60 watts, while the results are similar for the other power cap values. In addition, the enhanced strategy can further increase performance by up to to 8\% (Vacation with power cap set to 50 watts) and by 3.5\% on average. Differently from the lock-based case, both strategies of the proposed technique show an higher speed-up compared to the dual-phase technique by up to 21\% (ssca2 with power cap set to 50), and by 7.7\% and 10.7\% on average for the basic strategy and the enhanced strategy respectively. 

\subsection{Analysis of the Results}

As a first observation, results show that in various cases with locks, the error of our technique and of the dual-phase technique is very close to zero. This is due to the fact that, in our study, the scalability is limited for all applications when using locks. In these scenarios, the number of concurrent threads providing the higher throughput (that is selected by our technique and by the dual-phase technique) is low, thus the value of \textit{P-state} can be changed up to 0 while the power cap frontier is still far. This keeps the error very close to 0 since it is unlikely that the power cap is violated during the exploration procedure or due to workload variations.

The error is generally reduced with the enhanced strategy compared to the basic strategy, while also improving performance. This arises since the former is able to react along the time between two consecutive exploration procedures to the possible variations of the power consumption of the selected configurations, as discussed at the end of Section \ref{enanched_strategy}.

The speed-up with our technique is less than 1 only in one case, i.e. for Genome with transactions when the power cap value is equal to 50 watts. We note that Genome with transactions is highly scalable (see Figure \ref{throughput_analysis}). This leads both the baseline technique and our technique to select 20 as number of concurrent threads. As shown by the plot in Figure \ref{throughput_analysis}, the throughput of Genome with transactions is subject to noise when close to 20 threads . Also, we remark that our technique is able to react to workload variations also in terms of scalability. In this scenario, these factors cause lower performance with our technique due to the noise, which sometimes (wrongly) leads to temporarily selecting a less than optimal number of concurrent threads.

As expected, for lock-based synchronization the proposed technique technique shows similar results to the dual-phase technique since both techniques return the same configuration when the ascending part of the throughput curve is missing. For transaction-based synchronization, the highest speed-up improvements  over the dual-phase technique are obtained for Ssca2 and Genome which show a less than linear ascending part of the throughput curve for each fixed \textit{P-state} (Figure \ref{throughput_analysis}. As the most significant example, in Ssca2 the throughput slightly increases when increasing the number of threads from 6 to 15 which makes the dual-phase technique select a configuration with 15 threads. Differently, the proposed technique allocates the power budget more efficiently by selecting a configuration with a lower number of threads at an increased frequency. We should note that the benefits of the proposed technique over the dual-phase technique are not limited to applications that rely on transactional-based synchronization. Effectively, performance benefits should be obtained for any application with a throughput function that shows an ascending part followed by a descending, or only an ascending part that is less than linear. 

Overall, the results of our experiments study show that it is possible to achieve significant performance benefits by appropriately selecting the number of concurrent threads and CPU \textit{P-state} taking into consideration the scalability of the the specific multi-threaded application. As expected, compared to the baseline technique, the proposed solutions achieves the best results with poorly scalable applications, i.e. where contention is not minimal. Compared to the dual-phase technique, the exploration of the whole bi-dimensional space of configurations performed by the proposed technique can provide an appreciable improvement in performance for some applications, while achieving the same results for others. Finally, the enhanced strategy manages to further improve performance and reduce the power cap error over the basic strategy.

\section{Conclusions}
\label{conclusion}
In this work we introduced a novel power capping technique that, by jointly tuning the CPU performance state and the number of concurrent threads, improves the performance of multi-thread applications, specifically for applications that show less than linear scalability due to contention. Exploiting the results of a preliminary analysis, the proposed technique can return in linear time the optimal configuration which provides the highest performance between all configurations with power consumption lower than the power cap. We also present an enhanced strategy that by fluctuating between different configurations optimizes the dynamic allocation of the power budget, resulting in both increased performance and reduced power cap error. Compared to the baseline technique, that always assigns to the application the highest possible number of cores, our strategy provides an average speed-up of 1.48x, with individual test cases reaching up to 2.32x. Furthermore, we show that by exploring the overall bi-dimensional space of configuration, the proposed technique can improve performance by up to 21\% compared to techniques that tune the number of threads and the CPU performance state independently. 

\bibliographystyle{IEEEtran}



%

\end{document}